\documentclass[twocolumn,preprintnumbers,superscriptaddress,nofootinbib,aps,prd,floatfix]{revtex4}

\usepackage{lipsum}
\usepackage{footmisc,enumerate}
\usepackage{multirow}
\usepackage{subfigure}
\usepackage{amsmath}
\usepackage{graphicx}
\usepackage{hyperref}
\usepackage{mathrsfs}
\usepackage{amssymb}
\usepackage{bbm}
\usepackage{mathtools}
\usepackage{dsfont}
\usepackage{siunitx}
\usepackage{physics}
\usepackage{float}
\usepackage{enumitem}
\usepackage{cancel}
\hypersetup{colorlinks=true, citecolor=blue, urlcolor=blue, linkcolor=blue}
\usepackage[normalem]{ulem}

\newcommand{\diag}{\mathop{\mathrm{diag}}}

\begin{document}

\newcommand{\og}{\ensuremath{\tilde{O}_g}\xspace}
\newcommand{\ot}{\ensuremath{\tilde{O}_t}\xspace}

\preprint{}

\title{The Weinberg Angle and 5D RGE effects in a $SO(11)$ GUT theory}

\begin{abstract}
The Weinberg angle is an important parameter in Grand Unified Theories (GUT) as its size is crucially influenced
by the assumption of unification. In scenarios with different steps of symmetry breaking, in particular in models
that involve gauge-Higgs unification, the connection of the ultraviolet theory and the TeV scale-relevant, effective
Standard Model description is an important test of the models' validity. In this work, we consider a 6D gauge-Higgs
unification GUT scenario and explore the TeV scale-GUT relation using a detailed RGE analysis in the 4D and 5D regimes of the theory, including constraints
from LHC measurements. We show that such can be consistent with unification in the light of current constraints, while
the Weinberg angle likely translates into concrete conditions on the fermion sector in the higher dimensional setup.
\end{abstract}

\author{Christoph Englert} \email{christoph.englert@glasgow.ac.uk}
\affiliation{SUPA, School of Physics \& Astronomy, University of Glasgow, Glasgow G12 8QQ, UK\\[0.1cm]}
\author{David J.~Miller} \email{david.j.miller@glasgow.ac.uk}
\affiliation{SUPA, School of Physics \& Astronomy, University of Glasgow, Glasgow G12 8QQ, UK\\[0.1cm]}
\author{Dumitru Dan Smaranda} \email{d.smaranda.1@research.gla.ac.uk}
\affiliation{SUPA, School of Physics \& Astronomy, University of Glasgow, Glasgow G12 8QQ, UK\\[0.1cm]}


\maketitle

\section{Introduction}
\label{sec:Intro}
The interaction structure of the Standard Model of Particle Physics (SM) strongly suggests a mechanism of unification. On the one hand, Grand Unified Theories (GUTs) elegantly address questions related to fermion charge assignments in addition to a range of other shortcomings that are present in the SM. Along these lines a range of less traditional approaches to grand unification have been proposed recently (for a recent review see e.g.~\cite{Croon:2019kpe}). A scenario that we will focus on in this work is grand unification in the context of gauge-Higgs unification~\cite{Espinosa:1998re,Hall:2001zb,Burdman:2002se,Medina:2007hz,Hosotani:2004wv,Terazawa:1977hq,Lim:2007jv}. In particular, we will focus on the model of Refs.~\cite{Hosotani:2017edv,Hosotani:2017hmu}. As shown in Ref.~\cite{Englert:2019xhz}, this model is consistent with current LHC measurements with future LHC measurements being able to extend the currently observed sensitivity to exotic states to the multi-TeV range.

If a new state is discovered in the future, a question that will arise as part of the ensuing characterisation programme is its role as a potential harbinger of unification. Answers to this question
will be model-dependent but can be informed by theoretical consistency arguments. One of
these consistency arguments that is typically highlighted in GUT scenarios is the tree-level prediction
of the Weinberg angle
\begin{equation}
\label{eq:weinberg}
\sin^2\theta_W = {3\over 8}\,,
\end{equation}
as a consequence of an (intermediate) SU(5) unification~\cite{Georgi:1974sy,Georgi:1974yf,Marciano:1979yg}. In perturbative theories, reproducing
this value in the UV is critical to support the hypothesis of unification. The relation of Eq.~\eqref{eq:weinberg}
receives perturbative corrections that will modify its value in the UV as a function of the theories fundamental
input parameters. However, the dominant relation between UV and TeV scales
is captured in the renormalisation group running of $\sin^2\theta_W$, i.e. starting from the observed value at the
electroweak scale and including corrections from new particles becoming accessible we should approach
the relation of Eq.~\eqref{eq:weinberg} or discover the necessity of additional model constraints.

This is the focus of this work in the context of the aforementioned gauge-Higgs unification scenario of Refs.~\cite{Hosotani:2017edv,Hosotani:2017hmu}. We
perform a detailed renormalisation group equation (RGE) investigation of the 4D and 5D phases of the scenario with a particular focus on the weak mixing angle. In Sec.~\ref{sec:model} we briefly outline the model to make our work self-contained. In Sec.~\ref{sec:RGEeff} we lay out the RGE solving methods within the respective 4D and 5D formalisms and discuss their qualitative behaviour using a particular parameter benchmark scenario. In Sec.~\ref{sec:WeinbergRGE} we comment on the Weinberg angle at the GUT scale as a means to gauge unification in the considered theoretical framework. Sec.~\ref{sec:ResDiscConc} is devoted to a numerical RGE scan. Particular attention is given to the number of RGE-active fermion generations that can provide guidance for future model-building. Sec.~\ref{sec:Conc} offers conclusions.

\section{The model}
\label{sec:model}
The model of Refs.~\cite{Hosotani:2017edv,Hosotani:2017hmu} is a 6D space-time with hybrid (warped+flat) compactification and an $SO(11)$ gauge symmetry, described by a Randall-Sundrum--like metric~\cite{Randall:1999ee}
\begin{equation}
 ds^2 = e^{-2 \sigma (y)} (\eta_{\mu\nu} dx^\mu dx^\nu  + d w^2)  + dy^2\,,
\end{equation}
where $e^{-2 \sigma (y)}$ is the warp factor associated with the $y\in [0, L_5]$ direction, $w \in [0, 2\pi R_6]$ is an euclidean direction, and $\eta_{\mu\nu} = \diag(-1 , +1, +1, +1)$ is the 4D Minkowski space-time metric. A $\mathds{Z}_2$ transformation $(x^\mu, y, w) \rightarrow (x^\mu, -y, -w)$ results in a $\mathcal{M}_4 \times (T^2 / \mathds{Z}_2)$
orbifold with 5D branes $\mathcal{M}_4 \times S^1 $, at the fixed points $y= 0, L_5 $. We assume a compactification $M_\text{GUT}^{-1}\sim R_6\ll \left\{ \pi k/(z_L - 1)\right\}^{-1}$, where $k$ is the $\mathrm{AdS}_5$ curvature and $z_L = e^{k L_5}$, implying Kaluza-Klein (KK) mass scales of the 5th and 6th dimension $m_{\text{KK}_5} \ll m_{\text{KK}_6}\sim M_\text{GUT}$.
The matter content as well as its localisation on the orbifold fixed points is given in Tab.~\ref{table:ModelFields}.
%
\begin{table*}[!t]
  \centering
  \parbox{0.6\textwidth}{
  \begin{tabular}{|c|c|c|c|}
  \toprule
  Name              &    Field          &   $SO(11)$ rep.      &  Bulk/Brane  \\
  \colrule
  Gauge bosons      &  $A_M (x, y, w)$            & $\mathbf{55}$                  & 6D Bulk \\
  Dirac spinors           &  $\Psi^\alpha_\mathbf{32}(x, y, w)$  & $\mathbf{32}$         & 6D Bulk \\
  Dirac vectors      &  $\Psi_\mathbf{11}^\beta (x, y, w)$  & $\mathbf{11}$            & 6D Bulk \\
  Dirac vectors      &  $\Psi'^{\beta}_\mathbf{11}(x, y, w)$  & $\mathbf{11}$            & 6D Bulk \\
  Spinor scalar      &  $\Phi_\mathbf{32}(x, w)$  & $\mathbf{32}$            & 5D Brane at $y=0$\\
  Majorana spinor & $\chi_\mathbf{1}^\beta(x, w)$ & $\mathbf{1}$ & 5D Brane at $y=0$ \\
  \toprule
 \end{tabular}}
 \parbox{0.35\textwidth}{
 \vspace{1.2cm}
 \caption{Field content of the model of Refs.~\cite{Hosotani:2017edv,Hosotani:2017hmu}. The columns provide details of the fields content, their transformation properties under the $SO(11)$ gauge symmetry, and their localisations in the 6D setup. $\alpha, \beta$ are generational indices where $\alpha = 1, 2, 3, 4$ and $\beta = 1, 2, 3$.}
 \label{table:ModelFields}}
\end{table*}

\begin{figure}[!b]
 \begin{center}
  \includegraphics[width=1.0\columnwidth]{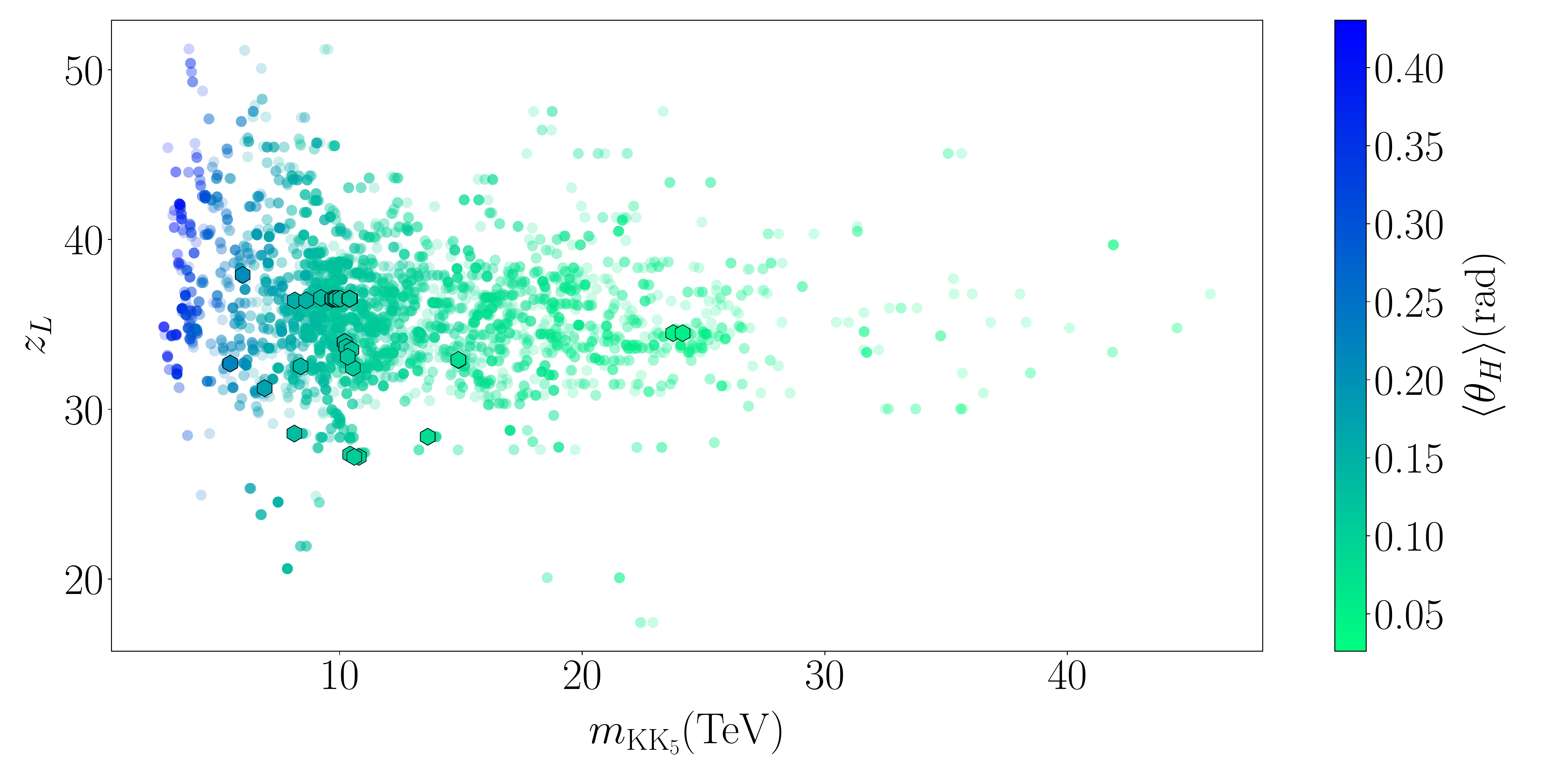}
  \caption{Scatter plot of representative parameter space points for the $SO(11)$ model as functions of the KK scale $m_{\text{KK}_5}$ and warp factor $z_L$. The color reflects the order parameter $\langle\theta_H\rangle$. Points highlighted as hexagons are points that are SM-like, i.e. they reproduce the SM in the low energy regime at the 95\% confidence level~\cite{Englert:2019xhz}. Faded points do not meet the 95\% confidence level criteria.}\label{fig:Res}
 \end{center}
\end{figure}

Symmetry breaking to Quantum Chromodynamics (QCD) and Electrodynamics (QED) proceeds in three stages: Firstly,
orbifolding with appropriate parity assignments~\cite{Scherk:1978ta,Scherk:1979zr} breaks $SO(11)\to G_\text{PS}= SU(4)_C \times SU(2)_L \times SU(2)_R$, the Pati-Salam~\cite{Pati:1974yy} group on the infrared (IR) brane at $y=L_5$. Secondly, 5D brane-localised interactions at $y=0$ of $\Phi_\mathbf{32}$ break $SO(11)\to SU(5)$
spontaneously, leading to a $SU(5) \cap G_\text{PS} = G_\text{SM} = SU(3)_C\times  SU(2)_L \times U(1)_Y$ zero mode spectrum in the gauge field KK decomposition.
Finally, below the 5D compactification scale (i.e. where a 4D description of the theory is appropriate), the Hosotani mechanism~\cite{Hosotani:1983xw, Hosotani:1988bm, Hosotani:1983vn} breaks $ SU(2)_L \times U(1)_Y \to U(1)_{\text{EM}}$ through a vacuum expectation value of a Wilson loop $\theta_H$ along the $y$ direction that carries the quantum numbers of the SM Higgs field. In addition to recreating the SM at the electroweak scale, the theory predicts KK towers for the $SO(11)$ gauge bosons and bulk matter fields in Tab.~\ref{table:ModelFields}. The masses of these modes are set by the various symmetry breaking stages and the two associated mass scales $m_{\text{KK}_5}, m_{\text{KK}_6}$.

For the purposes of exploring the model's parameter space, as done in e.g. \cite{Hosotani:2017edv}, we identify the Weinberg angle at the electroweak scale as $\sin^2\theta_W = 0.2312$. As shown in Ref.~\cite{Englert:2019xhz} the parameter region leading to an acceptable low energy phenomenology can be extended with adapted statistical sampling methods. This is highlighted in Fig.~\ref{fig:Res}, where we identify a parameter point as ``SM-like'' when it reproduces the SM at the $95\%$ confidence limit.\footnote{We refer the interested reader to Ref.~\cite{Englert:2019xhz} for details.}

\begin{figure}[!b]
 \begin{center}
   \includegraphics[width=1.0\columnwidth]{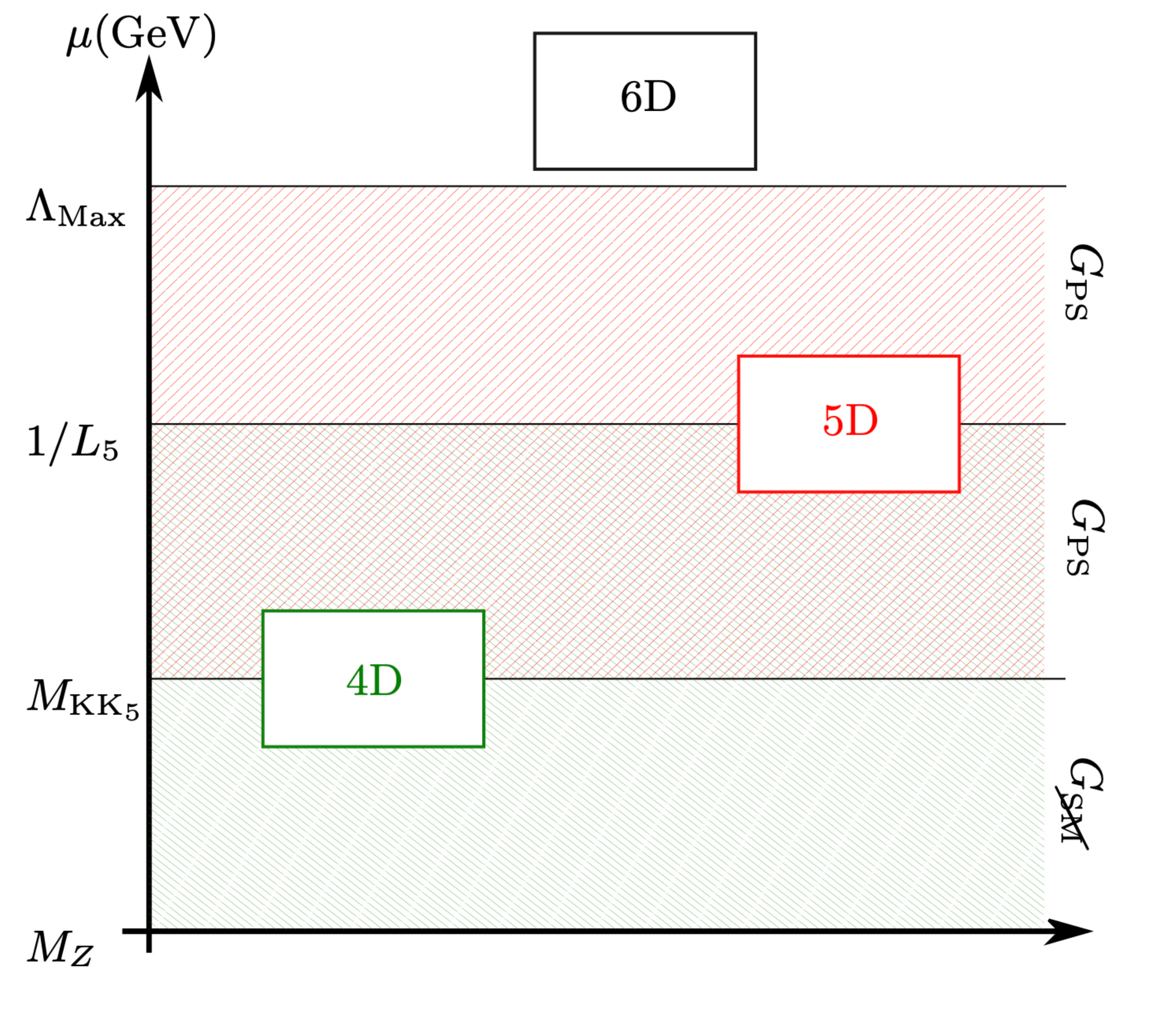}
  \caption{Tower of EFTs that approximate the UV 6D theory. The 4D description is valid within $[M_Z, M_{\text{KK}_5}]$ with $G_{\cancel{\text{SM}}} \equiv SU(3)_C \times U(1)_\text{EM}$ gauge symmetry and within $[M_{\text{KK}_5}, 1 / L_5 ]$ with $G_\text{PS}$ gauge symmetry. The 5D description is valid within $[M_{\text{KK}_5}, \Lambda_\text{Max}]$ with a  $G_\text{PS}$ gauge symmetry. Above $\Lambda_\text{Max}$ the full 6D description comes into effect.}\label{TowerOfTheories}
 \end{center}
\end{figure}

\section{RGE effects}\label{sec:RGEeff}
\subsection*{General remarks}
At the TeV scale the model is effectively the 4D SM and we evolve the parameters according to the 4D theory properties. This is admissible until we approach $M_{\text{KK}_5}$ where the 5D structure becomes apparent. At this stage we could continue using 4D RGE equations including the additional KK states that have non-trivial quantum numbers under the SM gauge group. Alternatively, one can directly work in a 5D approximation~\cite{Choi:2002ps} of the theory to obtain identical results, see Fig.~\ref{TowerOfTheories}. Above the $M_{\text{KK}_5}$ scale additional KK states of the 5D theory become accessible which correct the behaviour of the 5D running.

The 5D regime is determined by the Pati-Salam symmetry group together with the active KK states and thresholds. 6D compactification effects are not relevant in this context as we assume $M_{\text{KK}_5} \ll M_\text{GUT} \sim 1/ R_6$. Without a 6D RGE formalism, a complete evolution to the GUT scale in our one-loop analysis is not possible since there is a scale
\begin{equation}
 \Lambda_\text{Max} \sim \frac{16 \pi^2}{g_5^2} \ll M_\text{GUT}\,,
\end{equation}
which signifies a loss of perturbative control of the 5D regime before the unification scale. In this work, we opt to understand this scale as a lower bound on the GUT scale itself and use the difference of the Weinberg angle with respect to Eq.~\eqref{eq:weinberg} as a measure to gauge unification qualitatively.

The gauge-related states with masses $\mathcal{O}(M_{\text{KK}_5})$ relevant for our discussion are gauge fields that transform under the symmetries
\begin{equation}
 A_M \sim \begin{cases}
 G_\text{PS} / G_\text{SM} \\
 G_\text{SM} \\
 SO(5) / SO(4)
 \end{cases}\hspace{-0.2cm}.
\end{equation}
In the theory's 5D regime, these states have defined transformation properties under the Pati-Salam $G_\text{PS}$ symmetry.
The coset $SO(5) / SO(4)$ sector which transforms as $(1, \mathbf{2}, \mathbf{2})$ under $G_\text{PS}$ and eventually
triggers electroweak symmetry breaking via the Hosotani mechanism, induces corrections to the gauge couplings $g_{2L}, g_{2R}$.
Note that the $w$ gauge component KK states of $SO(5)/SO(4)$ obtain large masses via brane interactions (see~\cite{Hosotani:2017edv}) and are therefore not relevant for our discussion.

The fermionic matter content relevant in the same regime, is again characterised by symmetry properties under $G_\text{PS}$. States with masses $ \mathcal{O}(M_{\text{KK}_5})$ are given by
  \begin{align}
    \begin{split}
     & (\mathbf{4}, \mathbf{2}, 1)_{L,R}, \quad (\mathbf{4}, 1, \mathbf{2})_{L,R}, \\
     & ( \mathbf{6}, 1, 1)^{(+)}_{L,R},  \quad (\mathbf{6}, 1, 1)^{(- )}_{L,R}, \\
     & (1, \mathbf{2}, \mathbf{2})^{(+)}_{L,R}, \quad (1, \mathbf{2}, \mathbf{2})^{(-)}_{L,R}, \quad (1, 1, 1)^{(+)}_{L,R}, \quad  (1, 1, 1)^{(-)}_{L,R},
   \end{split}
   \label{MatterContent}
  \end{align}
which all originate from the $\Psi^\alpha_\mathbf{32}, \Psi_\mathbf{11}^\beta, \Psi_\mathbf{11}'^{\beta}$ bulk fields. The signs $\pm$ refer to parity assignments to guarantee 6D $SO(11)$ chiral anomaly cancellation, see Ref.~\cite{Hosotani:2017edv}.

We divide the full energy range in which the 5D EFT is valid (i.e. $[M_Z, \Lambda_\text{Max}]$) into two regions. The first region is given by the energy range in which the 5D EFT is well-approximated by its 4D EFT counterpart. This region's cut-off energy is dictated by the $M_{\text{KK}_5}$ mass threshold around where the gauge bosons of the Pati-Salam symmetry are resolved.
This corresponds to a scale given by the first non-zero mode of the photon tower.\footnote{
For warp factor choices $z_L > 10$ that yield realistic low energy spectra, the solutions for the first photon mode and the PS gauge bosons are almost degenerate.}
Thus, the first region is very well approximated by a $G_\text{SM}$ theory with additional matter states (that correspond to the $\theta_H$ shifted KK towers), which is valid between $[M_Z, M_{\text{KK}_5}]$. We describe the remaining energy range  $[M_{\text{KK}_5}, \Lambda_\text{Max}]$ in the 5D $G_\text{PS}$ formalism following \cite{Choi:2002ps}, where the cut-off represents the energy at which we lose perturbative control of the 5D theory, and the more fundamental 6D theory is required.
The tower of theories is schematically shown in Fig.~\ref{TowerOfTheories}.

We now turn to the discussion of the 4D evolution, which will provide the IR boundary conditions for the 5D theory.
We first fix our (electroweak) input parameters at $M_Z$ by setting $\alpha_{3C}, \alpha_\text{EM}, \sin \theta_W$
to their experimentally observed values~\cite{Chakraborty:2014aca, Mohr:2015ccw}
\begin{equation}
\begin{split}
 \alpha_{3C} &= 0.11822\,,\\ 
 \alpha_\text{EM}^{-1}& = 127.916\,, \\
  \sin^2 \theta_W &= 0.2312\,,
\end{split}
\end{equation}
where $\alpha_{3C}, \alpha_\text{EM}$ denote the strong and electric structure constants, respectively (we will discuss the impact of uncertainties on our results below). Subsequently, we then evolve $\alpha_{3C}, \alpha_\text{EM}, \sin \theta_W$ via the  $G_\text{SM}$ RGEs in the broken phase (using the formalism outlined in  \cite{Erler:2004in}) until we reach the energy scale at which a new KK state becomes available.

At this scale, we include new RGE contributions arising from resolved KK states until we reach $M_{\text{KK}_5}$, where we include threshold corrections $\lambda_i$ corresponding to integrating out
the heavy states corresponding to the $G_\text{PS} \rightarrow G_\text{SM}$ breaking (we do not include logarithmic threshold corrections arising from the matter fields.).

The 4D/5D matching requires the identification of coupling constants at the relevant scale. The electroweak couplings of the unbroken $SU(2)_L\times U(1)_Y$ phase, $ \alpha_{1Y}, \alpha_{2L}$ are related to their broken phase counterparts by\footnote{This is done in the 4D framework, and we have adopted the standard $3/5$ GUT normalisation for the hypercharge coupling.}
 \begin{equation}
 \begin{split}
   \frac{1}{\alpha_{1Y} (\mu)} \eval_{\mu = M_{\text{KK}_5} } &= \frac{3}{5} (1- \sin^2 \theta_W) \frac{1}{\alpha_\text{EM} (\mu)} \eval_{\mu = M_{\text{KK}_5} } \,, \\
   \frac{1}{\alpha_{2L} (\mu) } \eval_{\mu = M_{\text{KK}_5} } &= \sin^2 \theta_W \frac{1}{\alpha_\text{EM} (\mu)} \eval_{\mu = M_{\text{KK}_5} }\,.
   \end{split}
 \end{equation}
\begin{figure*}[!t]
 \begin{center}
  \includegraphics[scale = 0.43]{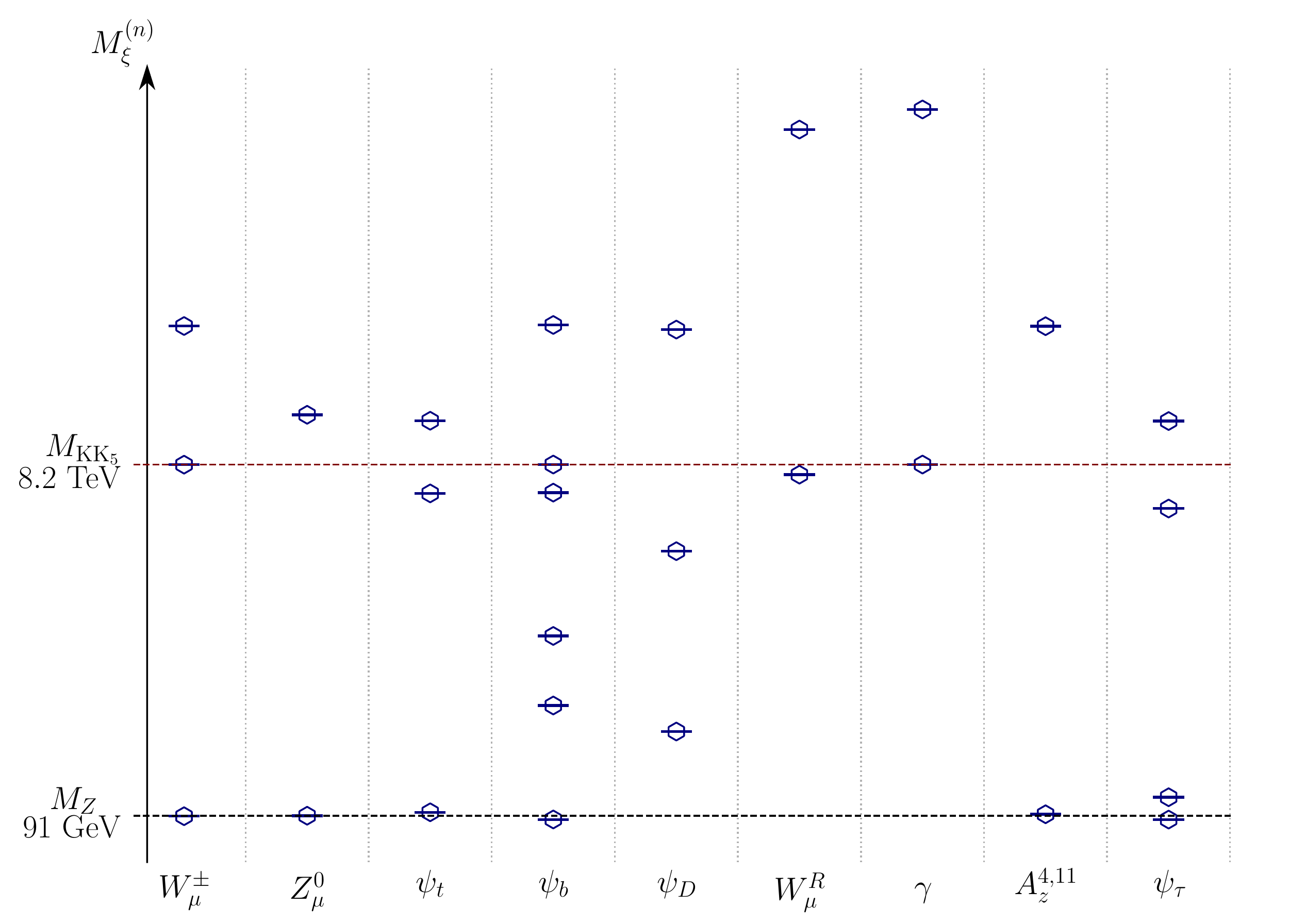}
  \caption{Tower of states from $M_Z$ to the next states at scales beyond $ M_{\text{KK}_5}$. The labels indicate the relevant fermion and boson fields, and their markers show the mass of the respective KK state.
  $W^{\pm}_\mu$ refer to the W boson tower,  $Z^{0}_\mu$  to the Z boson tower, $\psi_t$ denotes the top quark tower, $\psi_b$ is the bottom quark tower, $\psi_D$ is the ``dark fermion'' multiplet tower, $W^{R}_\mu$ is the Pati-Salam $SU(2)_R$ W boson tower, $\gamma_\mu$ is the photon tower,  $A^{4, 11}_z$ is the Higgs tower, and $\psi_{\tau}$ is the tau tower.
  }\label{KKTowerExplicit}
 \end{center}
\end{figure*}
%
With this we can now find the values of the Pati-Salam gauge couplings $\alpha_{4C}, \alpha_{2L}, \alpha_{2R}$ at the $M_{\text{KK}_5}$ scale
\begin{equation}
\begin{split}
\frac{1}{\alpha_{4C}} & = \frac{1}{\alpha_{3C}} + \frac{1 }{12 \pi} \,,\\
\frac{1}{\alpha_{2R}} & = \frac{5}{3} \frac{1}{\alpha_{1Y}} - \frac{2}{3} \frac{1}{\alpha_{3C}} + \frac{8}{45 \pi} \,,
\label{eq:paticoup}
\end{split}
\end{equation}
($\alpha_{2L}$ is already given as the coupling of the $SU(2)_L$ group). These
serve as the boundary conditions for the 5D theory, where
\begin{equation}
  g_{\mathrm{5D}} \sqrt{L_5} = g_{\mathrm{4D}}\eval_{\mu = M_{\text{KK}_5}}.
\end{equation}
We evolve the Pati-Salam couplings $\alpha_4, \alpha_{2L}, \alpha_{2R}$ within the 5D formalism described in Ref.~\cite{Choi:2002ps} in the energy range $[M_{\text{KK}_5}, \Lambda_\text{Max}]$. Using this running we then extract the coupling values and compare the Weinberg angle
\begin{multline}
  \sin^2 \theta_W (\mu)  = \Bigg(  \frac{1}{\alpha_{2L}\alpha_{4C}}  \Big( \alpha_{2L} \alpha_{4C} + \frac{2}{3} \alpha_{2L}\alpha_{2R} +  \\
  \alpha_{2R}\alpha_{4C} - \frac{5}{3} \alpha_{2L}\alpha_{2R} \alpha_{4C} \frac{8}{45 \pi} \Big) \Bigg)^{-1}\eval_{\mu} \label{eqn:WeinbergAnglePS}
\end{multline}
to its predicted GUT value.

Before we discuss the RGEs in detail below, it is instructive to define a reference point to guide our discussion. To get a qualitative understanding of how the KK thresholds modify the RG evolution of the theory, we consider the set of parameters from~\cite{Hosotani:2017edv}, which provide a SM-like physical mass spectrum
\begin{align}
  \begin{split}
    \mathcal{P}_\text{sample} :=  & \Big\{ k = 89130, z_L = 35, c_1 = 0, c_2 = -0.7,  \\
    &   c'_0 = 0.5224, \mu_1 = 11.18, \mu_{11} = 0.108,   \\
    & \tilde{\mu}_2 = 0.7091 , \mu'_{11} = 0.108 \Big\} \,.
  \end{split}\label{eqn:samplePoint}
\end{align}
$c_1, c_2, c'_0$ are the fermion bulk mass parameters along the warped direction, and $\mu_{1}, \tilde{\mu}_2, \mu_\textbf{11}, \mu'_\textbf{11}$ are couplings localised on the 5D brane at $y=0$ (for details see \cite{Hosotani:2017edv}).
This choice results in the tower of states shown in Fig.~\ref{KKTowerExplicit}, which we will use as a reference point in the following.

\subsection*{4D Approximation and RGEs}
By performing the RGE analysis in the broken phase, we evolve the QCD gauge coupling $g_3$, along with the electromagnetic coupling  $g_\text{EM}$, which in turn determines the Weinberg angle $\sin \theta_W$ RGE evolution via the matter content. To facilitate an unambiguous transition to the Pati-Salam phase we then proceed to relate the latter to the unbroken $U(1)_Y$ hypercharge and $SU(2)_L$ weak couplings.

The renormalisation group equations are expressed in terms of the gauge couplings $g_i$ as
\begin{equation}
  \mu \frac{\partial g_i}{ \partial \mu}  = \beta_i (g_i, \mu)\,, \qquad \frac{1}{\alpha_i} = \frac{4 \pi}{ g_i^2}\,,
\end{equation}
where $\beta_i$ are the beta coefficients arising from the group representations of the $SU(N)$ gauge group.
The QCD beta function $\beta_{g_3}$, has the generic form arising from a $SU(N)$ gauge theory \cite{Machacek:1984zw} with fermions and scalars in representations $F_i$ and $S_i$,
\begin{equation*}
 \beta_{g_3} = \frac{g_3^3}{(4\pi)^2} \left\{ -\frac{11}{3} C_2\left(SU(3)\right) + \frac{4}{3}\kappa S_2\left(F_i\right)  + \frac{1}{6} \eta S_2\left(S_i\right) \right\}
 \end{equation*}
where $C_2\left(G_i\right)$ is the quadratic Casimir of the group $G_i$, $S_2\left(F_i\right), S_2\left(S_i\right)$ are the Dynkin indices for the fermion/scalar representations, $\kappa = 1/2, 1$ for Weyl and Dirac fermions, respectively, and $\eta = 1, 2$ for real and complex scalar fields.

For the RGE runnings of the QED gauge coupling $g_\text{EM}$ and Weinberg angle $\sin \theta_W$, we use the formalism presented in Ref.~\cite{Erler:2004in}. The QED beta function is
\begin{equation*}
\beta_{g_\text{EM}} =  \frac{g_\text{EM}^3}{(4\pi)^2}  \frac{1}{6} \left\{ \sum_{i} N_i^c \gamma_i Q^2_i  \right\} \,,
\end{equation*}
where $N_i^c$ are the fermion colour factors, $Q_i$ are the EM charges and $\gamma_i = \{ -22, 8, 4, 2 \}$ correspond to gauge bosons, Dirac/chiral fermions and complex scalar fields.

We begin our RGE evolution at $M_Z \simeq \SI{91}{GeV}$. The QCD and QED couplings have beta function coefficients
\begin{equation}
\beta_{g_3} = -7 \frac{g_3^3}{(4\pi)^2}\, , \qquad{\beta_{ g_\text{EM} } } =  22 \frac{g_\text{EM}^3}{(4\pi)^2}\,,
\end{equation}
which are determined by the SM matter content and their $SU(3)_C$ and $U(1)_\text{EM}$ charges. As we evolve the couplings and encounter new states, the beta functions pick up new contributions. The additional contributions to the QCD beta function take the form
\begin{subequations}
\label{eq:betfu}
\begin{equation}
 \beta_{g_3} \rightarrow \beta_{g_3} +
 \begin{cases}
  -\frac{11}{3} C_2(SU(3))  \\
  +\frac{4}{3} \kappa S_2(F_i) \cdot N_G  \\
  +\frac{1}{6} \sum \eta S_2(S_i)
 \end{cases}
\end{equation}
depending on the nature of the state.
Analogously, for the QED beta function we have,
\begin{equation}
\beta_{g_\text{EM}} \rightarrow \beta_{g_\text{EM}} +
\begin{cases}
 -22 N_i^c \gamma_i Q_i^2  \\
 + 8 N_i^c \gamma_i Q_i^2 \cdot N_G \\
 +2 N_i^c \gamma_i Q_i^2
\end{cases} \hspace{-0.2cm}.
\end{equation}
\end{subequations}

In Eqs.~\eqref{eq:betfu}, we have introduced the $N_G$ factor in the fermionic contributions to account for the number of matter generations present in the model. In this paper we examine  the $N_G=1, 3$ cases. For $N_G = 3$ we assume that all three SM generations contribute and that the mass differences between the associated KK states is negligible for the non-zero modes. Similarly for $N_G = 1$ we assume that there is a mass separation mechanism between the third family and the other two which effectively decouples the non-zero states from the theory, leaving only the third as relevant, as in \cite{Hosotani:2017edv}. Comparing the different assumptions will point towards future model building directions (see below) in the light of expected unification.

With this framework in place we can now form a piecewise system of differential equations.
As shown in  \cite{Erler:2004in} the Weinberg angle's RGE running is fully determined by its experimental value at $M_Z$,
the matter content of the theory, and the running of $\alpha_\text{EM}$

 \begin{multline}
   \sin^2 \theta_W (\mu) = \frac{\alpha_\text{EM} (\mu)}{\alpha_\text{EM} (\mu_0)}  \sin^2 \theta_W (\mu_0)  \\
      + \frac{\sum_{i} N_i^c \gamma_i Q_i T_i }{\sum_{i} N_i^c \gamma_i Q^2_i } \left[1 - \frac{\alpha_\text{EM} (\mu)}{\alpha_\text{EM} (\mu_0)} \right]
    \,, \label{eqn:GaugeRel}
 \end{multline}
where $T_i$ is the third component of the weak isospin ($T_3 = +1/2$ for $u_i, \nu_i$, $T_3 = -1/2$ for $d_i, e_i$, $T_3 = \pm 1$ for $W^\pm$). The RGE running for the Weinberg angle starting at $M_Z$ is determined by the matter content of the SM and has a growth coefficient
\begin{equation}
  \frac{\sum_{i} N_i^c \gamma_i Q_i T_i }{\sum_{i} N_i^c \gamma_i Q^2_i } = -\frac{19}{22}.
\end{equation}
Therefore, using the numerical solution for $\alpha_\text{EM}$ we can now create an analogous piecewise solution for $\sin \theta_W$ based on the present matter content.
The fields' charges under $SU(3)_C \times U(1)_\text{EM}$, along with their $T_3$ values are given in Table \ref{table:RGEcharges}.

\begin{table}[!t]
  \centering
  \begin{tabular}{|c|c|c|c|}
  \toprule
  Name              & $SU(3)_C$ Charge              & $U(1)_\text{EM}$ Charge      &  $T_3$ \\
  \colrule
  Tau ($\tau$)      & 1                             & -1                           & -1/2 \\
  Bottom ($b$)      & 3                             & -1/3                         & -1/2 \\
  Top ($t$)         & 3                             & +2/3                         & +1/2 \\
  Neutrino ($\nu$)  & 1                             & 0                            & +1/2 \\
  $W^\pm$           & 1                              & $\pm 1$                        & $\pm 1$ \\
  Dark Fermion $\psi_D^{\nu}$          & 1          & 0                            & +1/2 \\
  Dark Fermion $\psi_D^{u}$            & 3          & +2/3                         & +1/2 \\
  Dark Fermion $\psi_D^{\overline{u}}$ & $\overline{3}$          & -2/3                         & -1/2 \\
  Dark Fermion $\psi_D^{e}$            & 1          & -1                           & +1/2 \\
  \toprule
 \end{tabular}
 \caption{Charge assignments for fields contributing to the RGE runnings.
 }
 \label{table:RGEcharges}
\end{table}

 \begin{figure}[t!]
  \begin{center}
   \includegraphics[width = 1.0\columnwidth]{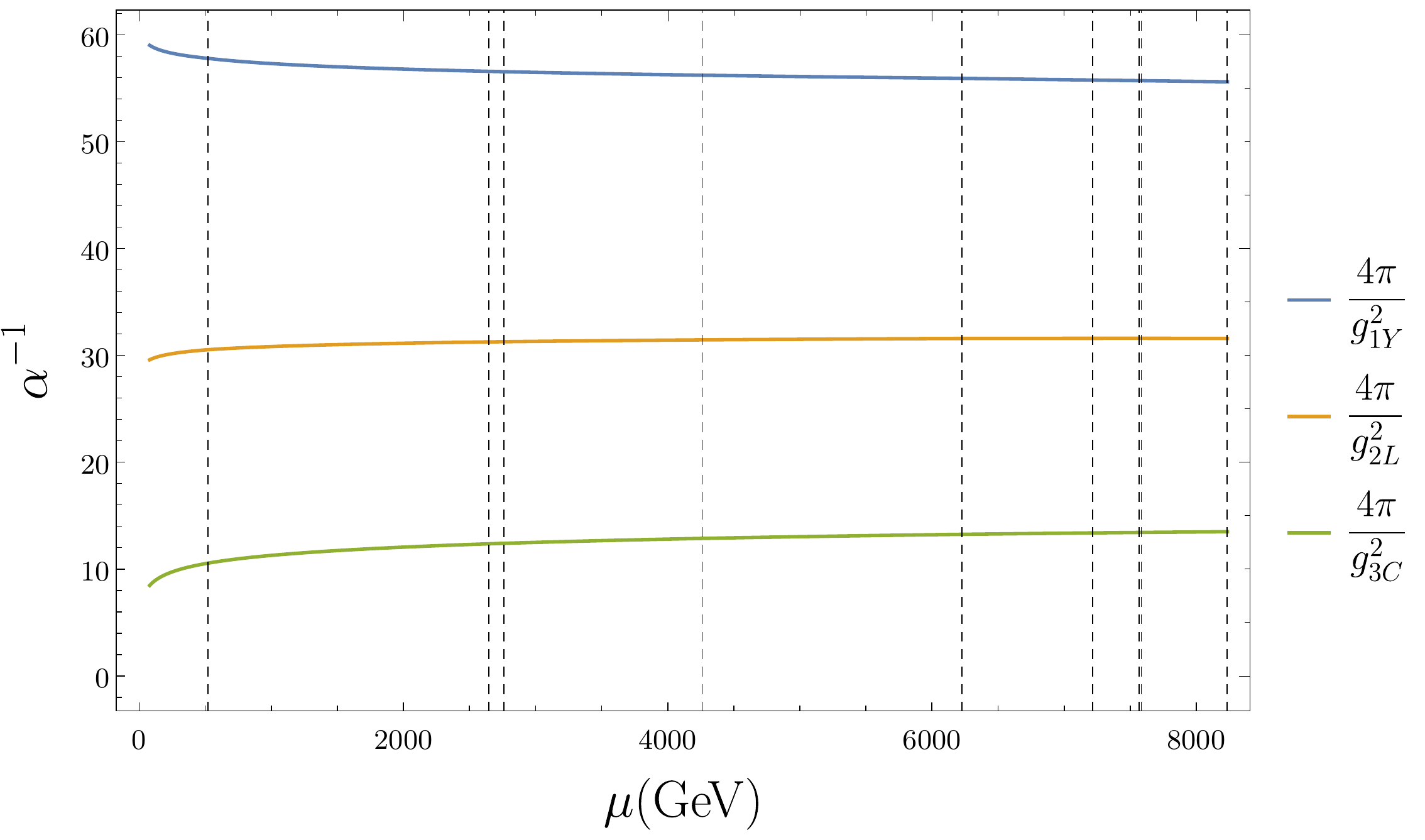}
   \caption{Piecewise RGE evolution for the SM couplings  $g_{3C}, g_{2L}, g_{1Y}$  with the different $\beta$ function changes at the multiple encountered KK states marked as dashed lines. ($M_{\text{KK}_5}$ itself is the furthest right dashed line.) Note that the piecewise forms for $g_{2L}, g_{1Y}$ are obtained via Eq.~\eqref{eq:weinbergc}. }\label{GSM_4DRGE}
  \end{center}
 \end{figure}

When we reach $M_{\text{KK}_5}$, we recover the hypercharge and weak couplings from the evolved values of $\alpha_\text{EM}$ and $\sin \theta_W$ via
   \begin{equation}
   \begin{split}
   \label{eq:weinbergc}
      \frac{1}{\alpha_{2L}(\mu)} & = \frac{1}{\alpha_\text{EM} (\mu)} \sin^2 \theta_W (\mu)\,, \\
      \frac{1}{\alpha_{1Y}(\mu)} & = \frac{3}{5} \frac{1}{\alpha_\text{EM}(\mu)} (1 - \sin^2 \theta_W(\mu))\,.
   \end{split}
   \end{equation}
Since $M_{\text{KK}_5}$ is the energy threshold at which the Pati-Salam states become available, we transition to the PS phase of the theory where we obtain the gauge couplings based on the symmetry breaking $SU(4)_C \times SU(2)_L \times SU(2)_R \rightarrow SU(3)_C \times SU(2)_L \times U(1)_Y$. This in turn provides us with the aforementioned 4D/5D boundary conditions of Eq.~\eqref{eq:paticoup}
evaluated at $M_{\text{KK}_5}$.
Following this procedure, the $G_\text{SM}$ gauge coupling running in the energy range $[M_Z,  M_{\text{KK}_5}]$ for the
spectrum of Fig.~\ref{KKTowerExplicit} is shown in Fig.~\ref{GSM_4DRGE}.

\subsection*{5D RGEs and cut-offs}
We now turn to the 5D running with the boundary conditions at $M_{\text{KK}_5}$ detailed above as input. The matter content in our approximated 5D theory was mentioned earlier in Eq.~\eqref{MatterContent}, in addition to the $(1, \mathbf{2}, \mathbf{2}) \sim SO(5)/SO(4)$ state.

The formalism in \cite{Choi:2002ps} specifies the 5D RGE running for generic 5D field parity assignments on a \mbox{$S^1/ \mathds{Z}_2 \times \mathds{Z}'_2$} orbifold. Since we started with a 6D theory defined on \mbox{$\mathcal{M}_4 \times T^2 / \mathds{Z}_2$}, the $S^1/ \mathds{Z}_2 \times \mathds{Z}'_2$ assignments arise from the orbifold assignments along the warped direction.
These assignments are tabled for fermions, gauge bosons and scalars in Tabs.~\ref{table:FermionParity},~\ref{table:GaugeParity}, and~\ref{table:GaugeScalarParity}, respectively.

\begin{table}[t!]
 \centering
 \begin{tabular}{c}
  \begin{tabular}{ | c | c | c |}
  \toprule
  $G_\text{PS}$ rep.              & Parent Field              & $(\mathds{Z}_2 , \mathds{Z}'_2)$ \\
  \colrule
  $(\mathbf{4}, \mathbf{2}, 1)_L$ & $\Psi^\alpha_\mathbf{32}$ & $(+,+)$                          \\
  $(\mathbf{4}, \mathbf{2}, 1)_R$ & $\Psi^\alpha_\mathbf{32}$ & $(-,-)$                          \\
  $(\mathbf{4}, 1, \mathbf{2})_R$ & $\Psi^\alpha_\mathbf{32}$ & $(+,+)$                          \\
  $(\mathbf{4}, 1, \mathbf{2})_L$ & $\Psi^\alpha_\mathbf{32}$ & $(-,-)$                          \\
  \toprule
 \end{tabular}
 \\ \\
 \begin{tabular}{ | c | c | c |}
  \toprule
  $G_\text{PS}$ rep.              & Parent Field         & $(\mathds{Z}_2 , \mathds{Z}'_2)$ \\
  \colrule
  $(\mathbf{4}, \mathbf{2}, 1)_L$ & $\Psi^4_\mathbf{32}$ & $(+,-)$                          \\
  $(\mathbf{4}, \mathbf{2}, 1)_R$ & $\Psi^4_\mathbf{32}$ & $(-,+)$                          \\
  $(\mathbf{4}, 1, \mathbf{2})_L$ & $\Psi^4_\mathbf{32}$ & $(-,+)$                          \\
  $(\mathbf{4}, 1, \mathbf{2})_R$ & $\Psi^4_\mathbf{32}$ & $(+,-)$                          \\
  \toprule
 \end{tabular}
 \\ \\
 \begin{tabular}{| c | c | c |}
  \toprule
  $G_\text{PS}$ rep.           & Parent Field             & $(\mathds{Z}_2 , \mathds{Z}'_2)$ \\
  \colrule
  $(\mathbf{6}, 1, 1)^{(+)}_R$ & $\Psi^\beta_\mathbf{11}$ & $(+,+)$                          \\
  $(\mathbf{6}, 1, 1)^{(-)}_L$ & $\Psi^\beta_\mathbf{11}$ & $(+,+)$                          \\
  $(\mathbf{6}, 1, 1)^{(+)}_L$ & $\Psi^\beta_\mathbf{11}$ & $(-,-)$                          \\
  $(\mathbf{6}, 1, 1)^{(-)}_R$ & $\Psi^\beta_\mathbf{11}$ & $(-,-)$                          \\
  \toprule
 \end{tabular}
 \\ \\
 \begin{tabular}{| c | c | c |}
  \toprule
  $G_\text{PS}$ rep.                          & Parent Field                & $(\mathds{Z}_2 , \mathds{Z}'_2)$ \\
  \colrule
  $(1, \mathbf{2}, \mathbf{2})^{(+,-)}_{L,R}$ & $\Psi^{\beta'}_\mathbf{11}$ & $(+,+)$                          \\
  $(1, 1, 1)^{(+,-)}_{R,L}$                   & $\Psi^{\beta'}_\mathbf{11}$ & $(+,+)$                          \\
  $(1, \mathbf{2}, \mathbf{2})^{(+,-)}_{R,L}$ & $\Psi^{\beta'}_\mathbf{11}$ & $(-,-)$                          \\
  $(1, 1, 1)^{(+,-)}_{L ,R}$                  & $\Psi^{\beta'}_\mathbf{11}$ & $(-,-)$                          \\
  \toprule
 \end{tabular}
 \end{tabular}
 \caption{Fermion Parity Assignments Under $S^1 / \mathds{Z}_2 \times \mathds{Z}'_2$.}
 \label{table:FermionParity}
\end{table}
\begin{table}[t!]
 \centering
 \begin{tabular}{c}
  \begin{tabular}{| c | c | c |}
  \toprule
  $G_\text{SM}$ rep.   & Parent Field            & $(\mathds{Z}_2 , \mathds{Z}'_2)$ \\
  \colrule
  $(1, \mathbf{3}, 0)$ & $A_\mu \in G_\text{SM}$ & $(+,+)$                          \\
  $(\mathbf{8}, 1, 0)$ & $A_\mu \in G_\text{SM}$ & $(+,+)$                          \\
  $(1, 1, 0)$          & $A_\mu \in G_\text{SM}$ & $(+,+)$                          \\
  \toprule
 \end{tabular}
 \\ \\
 \begin{tabular}{| c | c | c |}
  \toprule
  $G_\text{SM}$ rep.                             & Parent Field                           & $(\mathds{Z}_2 , \mathds{Z}'_2)$ \\
  \colrule
  $(\mathbf{3}, 1, 0 )$           & $A_\mu \in G_\text{PS} /  G_\text{SM}$ & $(-,+)$            \\
  $(\overline{\mathbf{3}}, 1, 0)$ & $A_\mu \in G_\text{PS} /  G_\text{SM}$ & $(-,+)$            \\
  \toprule
 \end{tabular}
 \\ \\
  \begin{tabular}{| c | c | c |}
   \toprule
   $G_\text{PS}$ rep.                                   & Parent Field                           & $(\mathds{Z}_2 , \mathds{Z}'_2)$ \\
   \colrule
   $(1, \mathbf{2}, \mathbf{2})$                        & $A^{a, 11}_\mu \in SO(5) / SO(4)$      & $(-,-)$                          \\
   $(1, 1, \mathbf{3}) $ & $W^\pm_R, Z_R \in G_\text{PS} /  G_\text{SM}$ & $(-,+)$                          \\
   \toprule
  \end{tabular}
 \end{tabular}
 \caption{Gauge boson parity assignment under \mbox{$S^1 / \mathds{Z}_2 \times \mathds{Z}'_2$}. Note that we have to treat the $G_\text{PS}$, and $G_\text{SM}$ representations separately due to the mixed parity assignments in the full 6D model.}
 \label{table:GaugeParity}
\end{table}
\begin{table}[ht!]
 \centering
 \begin{tabular}{c}
  \begin{tabular}{| c | c | c |}
  \toprule
  $G_\text{PS}$ rep.    & Parent Field          & $(\mathds{Z}_2 , \mathds{Z}'_2)$ \\
  \colrule
  $(\mathbf{15}, 1, 1 )$ & $A_y \in G_\text{PS}$ & $(-,-)$                          \\
  $(1, 1, \mathbf{3})$  & $A_y \in G_\text{PS}$ & $(-,-)$                          \\
  $(1, \mathbf{3}, 1)$  & $A_y \in G_\text{PS}$ & $(-,-)$                          \\
  \toprule
 \end{tabular}
 \\ \\
 \begin{tabular}{| c | c | c |}
  \toprule
  $G_\text{PS}$ rep.    & Parent Field            & $(\mathds{Z}_2 , \mathds{Z}'_2)$ \\
  \colrule
  $(\mathbf{15}, 1, 1)$ & $A_w \in G_\text{PS}$ & $(-,-)$                          \\
  $(1, 1, \mathbf{3})$  & $A_w \in G_\text{PS}$ & $(-,-)$                          \\
  $(1, \mathbf{3}, 1)$  & $A_w \in G_\text{PS}$ & $(-,-)$                          \\
  \toprule
 \end{tabular}
 \\ \\
  \begin{tabular}{| c | c | c |}
   \toprule
   $G_\text{PS}$ rep.            & Parent Field                    & $(\mathds{Z}_2 , \mathds{Z}'_2)$ \\
   \colrule
   $(1, \mathbf{2}, \mathbf{2})$ & $A^{4, 11}_y \in SO(5) / SO(4)$ & $(+,+)$                          \\
   \toprule
  \end{tabular}
 \end{tabular}
 \caption{Scalar parity assignment under $S^1 / \mathds{Z}_2 \times \mathds{Z}'_2$. In the 5D RGE formalism they are treated as scalars originating from either the gauge boson projections or as remnants from the 6D approximation. }
 \label{table:GaugeScalarParity}
\end{table}

The 5D RGEs take the generic form~\cite{Choi:2002ps}
\begin{equation}
 \frac{1}{g^2_a (\mu)} =  \frac{\pi L_5}{g^2_{a_{\mathrm{5D}}} \left( \Lambda_\text{Max} \right) } \color{black} + \frac{1}{8\pi^2} \sum_\xi \overline{\Delta}_a \left( \xi; \mu , \ln \Lambda_\text{Max} \right),
\end{equation}
where $g_a$ is the 4D gauge coupling  corresponding to the respective gauge group in $SU(4)_C \times SU(2)_L \times SU(2)_R$ (where by $a$ we denote $4C, 2L, 2R$), $g^2_{a_{\mathrm{5D}}}$ is the squared 5D gauge coupling (which has mass dimension $M^{-1}$).
$\overline{\Delta}_a$ (see Appendix \ref{appendix:5DRGEs}) are denote the one loop corrections due to the theory's field content labelled with $\xi \in \left\{ \phi, \psi, A_\mu  \right\}$ for scalars, fermions and gauge bosons.
$\overline{\Delta}_N (\xi)$ for a gauge group $SU(N)$ and a field $\xi$ are given in Ref.~\cite{Choi:2002ps} and reproduced in the appendix for completeness.

\begin{figure}[t!]
  \begin{center}
   \includegraphics[width = 1.0\columnwidth]{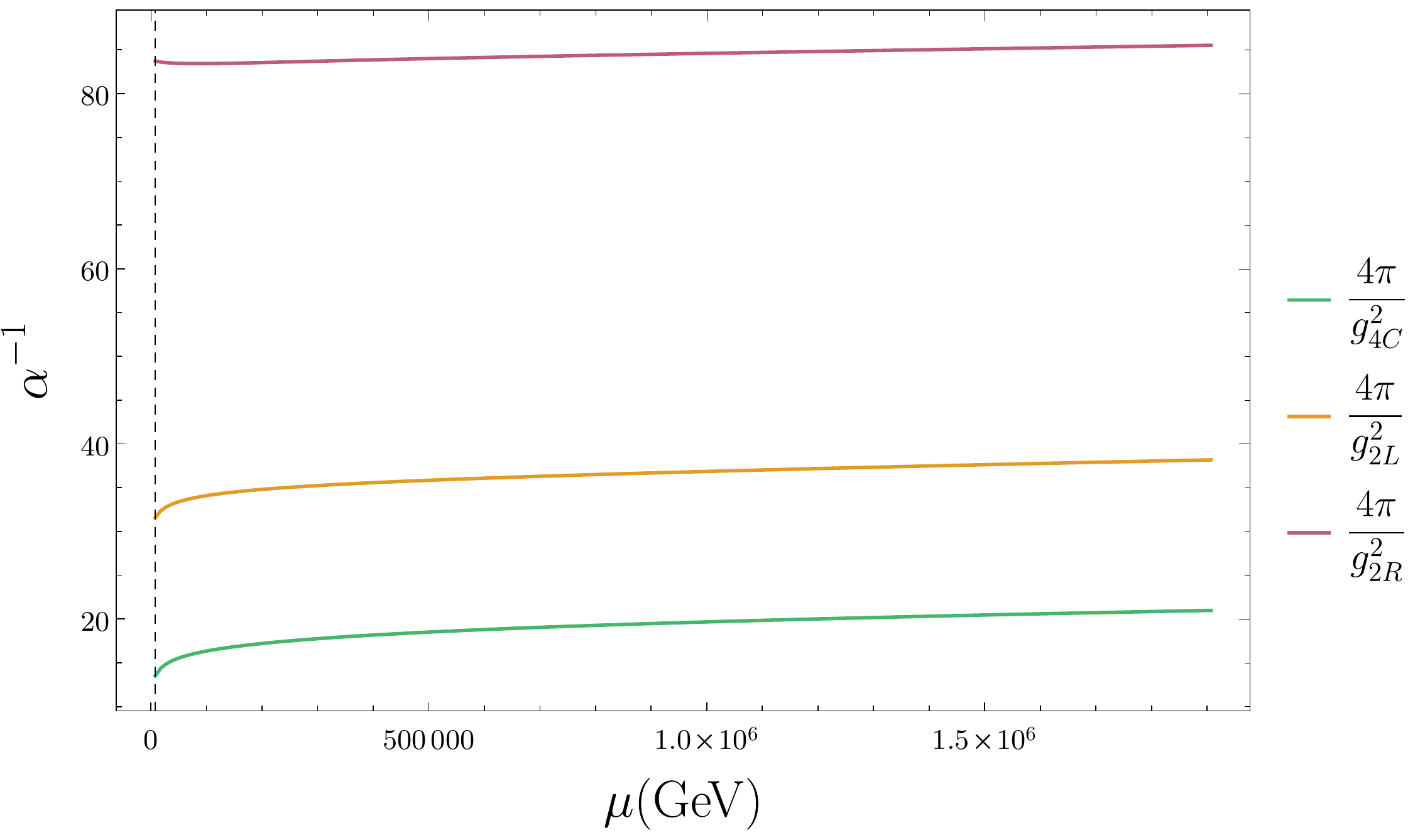}
   \caption{Effective 4D $SU(4)_C\times SU(2)_L \times SU(2)_R$ gauge couplings obtained via the 5D Pati-Salam approximation, using the evolved coupling values originating from the 4D formalism. The dotted line corresponds to the $M_{\text{KK}_5}$ threshold at which we start our 5D runnigs.}\label{protoRunning}
  \end{center}
 \end{figure}

We can define a cut-off $\Lambda_\text{Max}$ as the scale at which we lose perturbative control of the 5D theory,
\begin{equation}
 \Lambda_\text{Max} \simeq \frac{16 \pi^2}{ g^2_{a_{\mathrm{5D}} } (\Lambda_\text{Max}) }\,.
\end{equation}
This is the scale where the formal expansion parameter becomes too large (see Ref.~\cite{Sundrum:2005jf}) to deliver reliable results within the context of our leading order RGE analysis. To get a numerical estimate for $\Lambda_\text{Max}$ we can use the RGEs evaluated at $M_{\text{KK}_5}$, i.e.
\begin{multline}
  \frac{1}{g^2_a \left( M_{\text{KK}_5} \right) }  =  \frac{\pi L_5}{ g^2_{a_{\mathrm{5D}}} \left( \Lambda_\text{Max} \right) } \\+ \frac{1}{8\pi^2} \sum_\xi \overline{\Delta}_a \left( \xi; M_{\text{KK}_5} , \ln \Lambda_\text{Max} \right) \\
                                                  \equiv C^a_5 \left( \Lambda_\text{Max} \right) + \frac{1}{8\pi^2} \sum_\xi \overline{\Delta}_a \left( \xi; M_{\text{KK}_5} , \ln \Lambda_\text{Max} \right) .
\label{eqn:GaugeCt5D}
\end{multline}
This is an implicit equation for our unknown 5D gauge coupling at the cut-off scale.
To find the unknown dimensionless $C^a_5$ (and scale $\Lambda_\text{Max}$), we can recast the above as a functional equation and solve it numerically for $C^a_5$.
More specifically we can recast $\Lambda_\text{Max}$ as
\begin{equation}
\Lambda_\text{Max}
= \frac{16 \pi}{L_5} C^a_5\, ,
\end{equation}
which then provides us with the functional form when substituted into Eq.~\eqref{eqn:GaugeCt5D},
\begin{equation}
C^a_5 = \frac{1}{g^2_a \left( M_{\text{KK}_5} \right) } - \frac{1}{8\pi^2} \sum_\xi \overline{\Delta}_a \left( \xi; M_{\text{KK}_5} , \ln \left( \frac{16 \pi}{L_5} C^a_5 \right) \right).
\end{equation}
Solving this equation numerically yields cut-off scales for each of the gauge couplings $\Lambda_\text{Max}^{4C}, \Lambda_\text{Max}^{2L}, \Lambda_\text{Max}^{2R}$. For the remainder of this paper we will refer to the smallest of the three when discussing the cut-off of the theory where a more fundamental 6D theory should come into effect
\begin{equation}
\Lambda_\text{Max} = \min \left \{ \Lambda_\text{Max}^{4C}, \Lambda_\text{Max}^{2L}, \Lambda_\text{Max}^{2R} \right\} \,.
\end{equation}
The running in the 5D regime for our sample point is shown in Fig.~\ref{protoRunning}.

\section{Weinberg Angle: $SU(5)$ prediction vs running}
\label{sec:WeinbergRGE}
We can now turn to an analysis of the RGE-corrected Weinberg angle.
Switching from the broken $SU(3)_C \times U(1)_\text{EM}$ phase, to the $G_\text{SM}$ phase, the Weinberg angle $\sin \theta_W$ and the electromagnetic fine structure constant $\alpha_\text{EM}$, determine the weak and hypercharge couplings according to Eq.~\eqref{eq:weinbergc}.
Similarly the $G_\text{SM}$ couplings are related to the $G_\text{PS}$ ones as expressed in Eq.~\eqref{eq:paticoup},
leading to the Weinberg angle expression of Eq.~\eqref{eqn:WeinbergAnglePS}.
At the unification scale, i.e. the energy at which the first non-zero GUT KK state becomes available $m_{\text{KK}_6} \sim 1/ (2 \pi R_6)$, we can write a series of identities between the 4D, 5D, 6D couplings based on the principle that there is only one fundamental gauge coupling.

Before gauge symmetry breaking, the 5D and 4D equivalent $SO(11)$ couplings at the 5D Planck and IR branes are related to the 6D gauge coupling by,
 \begin{align*}
   \alpha^{SO(11)}_{\mathrm{6D}} & = \frac{\alpha^{SO(11)- \text{IR}}_{\mathrm{5D}} }{ 2 \pi R_6 } = \frac{\alpha^{SO(11)- \text{Pl}}_{\mathrm{5D}} }{ 2 \pi R_6 } \\
                        & = \frac{\alpha^{SO(11)- \text{IR}}_{\mathrm{4D}} }{ 2 \pi R_6 L_5 } = \frac{\alpha^{SO(11)- \text{Pl}}_{\mathrm{4D}} }{ 2 \pi R_6 L_5 } \,.
 \end{align*}
 On the Planck brane the gauge symmetry is broken down to  $SU(5)$ via the vacuum expectation value (VEV) $\langle \Phi_\mathbf{32} \rangle$. In terms of the equivalent 4D gauge couplings, the identification at 1-loop is equivalent to \cite{Hall:1980kf,Babu:2015bna}
 \begin{multline}
   \frac{1}{\alpha^{SU(5) - \text{Pl}}_{\mathrm{4D}}} = \frac{1}{\alpha^{SO(11) - \text{Pl}}_{\mathrm{4D}}}
   \\- \frac{1}{12\pi}\left[ C_2(SO(11)) - C_2(SU(5)) \right].
 \end{multline}
 Recasting this in terms of the 6D coupling, we have
 \begin{equation}
    \frac{1}{\alpha^{SU(5) - \text{Pl}}_{\mathrm{4D}}} = \left\{ \frac{1}{\alpha^{SO(11)}_{\mathrm{6D}}} -  2 \pi  R_6 L_5 \frac{\lambda_{11\rightarrow 5}}{12\pi} \right\} \frac{1}{2 \pi  R_6 L_5}\,, \label{eqn:SU5gauge}
 \end{equation}
 where $\lambda_{11\rightarrow 5} = \left[ C_2(SO(11)) - C_2(SU(5)) \right]$.
Similarly, on the IR brane we break $SO(11)\rightarrow SU(4)_C \times SU(2)_L \times SU(2)_R$ via
boundary conditions, which produce the gauge identifications at 1 loop,
 \begin{align*}
   \frac{1}{\alpha^{SU(4)_C - \text{IR}}_{\mathrm{4D}}} = \frac{1}{\alpha^{SO(11) - \text{Pl}}_{\mathrm{4D}}} - \frac{\lambda_{11\rightarrow 4}}{12\pi}\, ,\\
   \frac{1}{\alpha^{SU(2)_L - \text{IR}}_{\mathrm{4D}}} = \frac{1}{\alpha^{SO(11) - \text{Pl}}_{\mathrm{4D}}} - \frac{\lambda_{11\rightarrow 2}}{12\pi}\, ,\\
   \frac{1}{\alpha^{SU(2)_R - \text{IR}}_{\mathrm{4D}}} = \frac{1}{\alpha^{SO(11) - \text{Pl}}_{\mathrm{4D}}} - \frac{\lambda_{11\rightarrow 2}}{12\pi} \,,
 \end{align*}
 where $\lambda_{11\rightarrow 4} = C_2(SO(11)) - C_2(SU(4)) , \lambda_{11\rightarrow 2} = C_2(SO(11)) - C_2(SU(2)) $. In terms of the 6D couplings this means,
 \begin{equation}
 \begin{split}
   \frac{1}{\alpha^{SU(4)_C - \text{IR}}_{\mathrm{4D}}} =\left\{ \frac{1}{\alpha^{SO(11) - \text{Pl}}_{\mathrm{6D}}} - 2\pi R_6 L_5 \frac{\lambda_{11\rightarrow 4}}{12\pi} \right\} \frac{1}{ 2 \pi R_6 L_5 }\,, \\
   \frac{1}{\alpha^{SU(2)_L - \text{IR}}_{\mathrm{4D}}} =\left\{ \frac{1}{\alpha^{SO(11) - \text{Pl}}_{\mathrm{6D}}} - 2\pi R_6 L_5 \frac{\lambda_{11\rightarrow 2}}{12\pi} \right\} \frac{1}{ 2 \pi R_6 L_5 }\,,  \label{eqn:PatiSalamGauge}  \\
   \frac{1}{\alpha^{SU(2)_R - \text{IR}}_{\mathrm{4D}}} =\left\{ \frac{1}{\alpha^{SO(11) - \text{Pl}}_{\mathrm{6D}}} - 2\pi R_6 L_5 \frac{\lambda_{11\rightarrow 2}}{12\pi} \right\} \frac{1}{ 2 \pi R_6 L_5 }\,,
    \end{split}
 \end{equation}

Ignoring the Casimir terms for a moment to keep the discussion transparent, at the unification scale, instead of the Eqs.~\eqref{eqn:SU5gauge}, \eqref{eqn:PatiSalamGauge}, we have
 \begin{equation}
 \begin{split}
   \frac{1}{\alpha^{SU(4)_C - \text{IR}}_{\mathrm{4D}}} & = \frac{1}{\alpha^{SU(2)_L - \text{IR}}_{\mathrm{4D}}} = \frac{1}{\alpha^{SU(2)_R - \text{IR}}_{\mathrm{4D}}} \\
          &= \frac{1}{\alpha^{SU(5) - \text{Pl}}_{\mathrm{4D}}} = \frac{1}{\alpha^{SO(11) - \text{Pl}}_{\mathrm{6D}}} \frac{1}{ 2 \pi R_6 L_5 }\,.
 \end{split}
 \end{equation}
When combined with the expression for the Weinberg angle in the Pati-Salam phase, Eq.~\eqref{eqn:WeinbergAnglePS}, these relations lead to the expected
 \begin{equation}
    \label{eq:weinuv}
   \sin^2 \theta_W (\mu)  \eval_{\mu = (2\pi R_6)^{-1} }= \frac{1}{\frac{2}{3}+ 1 +1} = \frac{3}{8}\,.
 \end{equation}
In essence, this is the $SU(5)$ prediction translated from the Planck brane to the IR brane.\footnote{
The scale of $SU(5)$ breaking is dictated by $\langle \Phi_\mathbf{32}\rangle \sim R_6^{-1}$, which is localised on the UV brane $y=0$, i.e. the scale in Eq.~\eqref{eq:weinuv} is consistent.} Again, we emphasise that this scale is not accessible within our 5D formalism, but we can infer some useful conclusions depending on the values of the RGE runnings at $\Lambda_\mathrm{Max}$, as we will see in Sec.~\ref{sec:ResDiscConc}.

\begin{figure}[h!]
  \begin{center}
   \includegraphics[width = 1. \columnwidth]{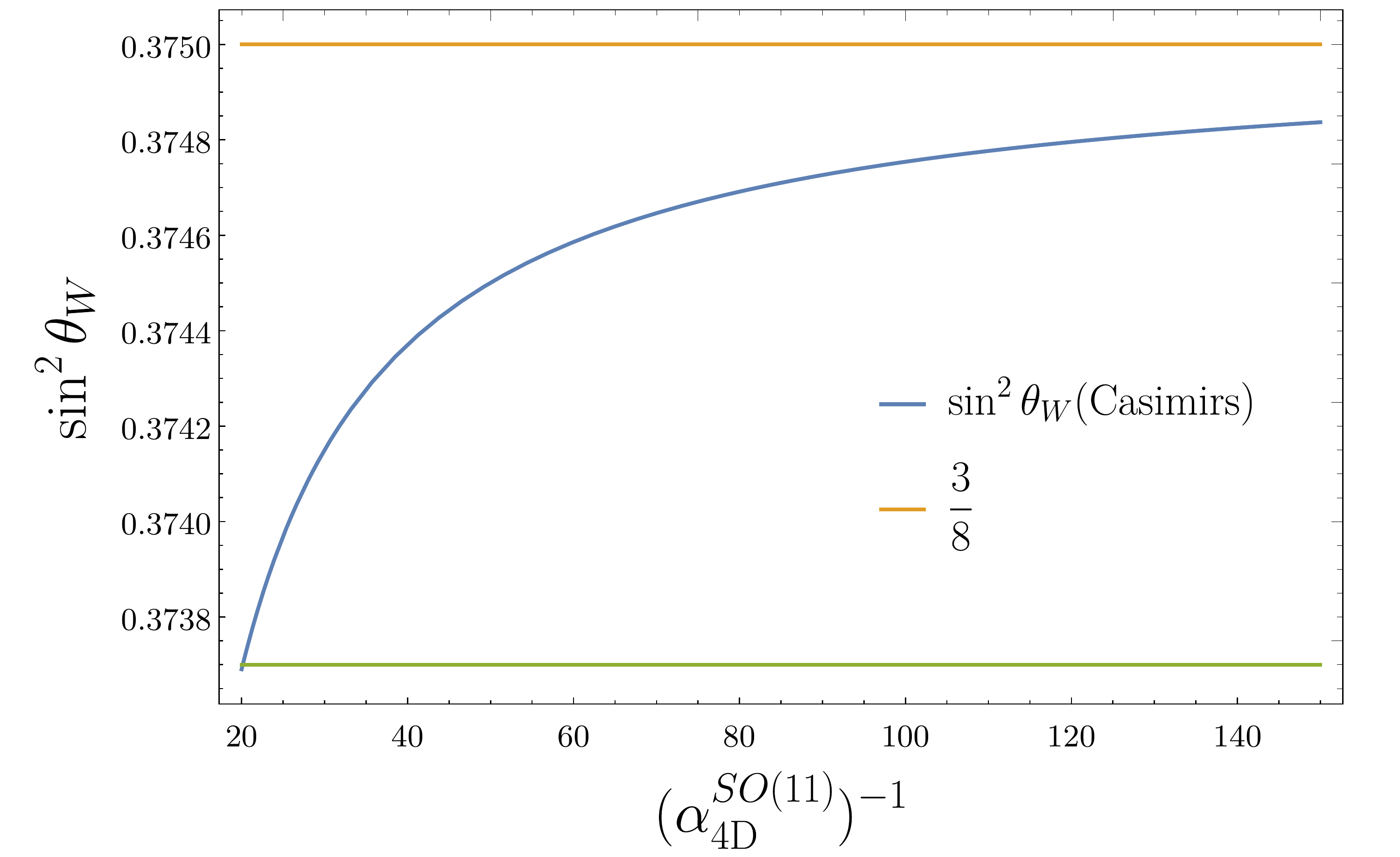}
   \caption{Numerical impact of the Casimir correction (blue line) as a function of the unknown inverse unified coupling $(\alpha_\mathrm{\mathrm{4D}}^{SO(11)})^{-1}$.
   The green line represents the low bound for the $\sim 0.4 \%$ deviation occurring at $(\alpha_\mathrm{\mathrm{4D}}^{SO(11)})^{-1} \simeq 20$.
   The orange line represents the GUT hypothesis 3/8. The smaller $\alpha$, the less impact the Casimir corrections have on the prediction as they weighted by $\alpha_\mathrm{\mathrm{4D}}^{SO(11)}$.
   }
 \label{figure:WeinCass}
  \end{center}
 \end{figure}

Including the Casimir corrections, we find the slightly modified relation
 \begin{equation}
   \sin^2 \theta_W (\mu)= \frac{36 - 18 \pi \alpha^{SO(11)}_{\mathrm{4D}}}{96 - \dfrac{1}{\pi} 20 \alpha^{SO(11)}_{\mathrm{4D}} - 44 \pi \alpha^{SO(11)}_{\mathrm{4D}}} \Biggr\rvert_{\mu = (2\pi R_6)^{-1} }.
 \end{equation}
 Since the Casimir-corrected Weinberg angle requires a value for the $SO(11)$ 4D equivalent gauge coupling, we examine the possible deviation from the $3/8$ GUT prediction as a function of the possible values of $\alpha^{SO(11)}_{\mathrm{4D}}$, as shown in Fig.~\ref{figure:WeinCass}. For reasonable $\alpha^{SO(11)}_{\mathrm{4D}}$ coupling values (e.g. Ref~\cite{Babu:2015bna}) we see that deviations arising from the Casimir-corrected values amount to $\lesssim - 0.0013$, see Fig.~\ref{figure:WeinCass}. Since this $\sim 0.4 \%$ deviation is negligible, we can safely ignore the Casimir contributions in the following without qualitatively changing our results.

\begin{figure*}[!t]
 \begin{center}
 \subfigure[\label{fig:SMabsDeV1Gena}]{\includegraphics[width=1.0\columnwidth]{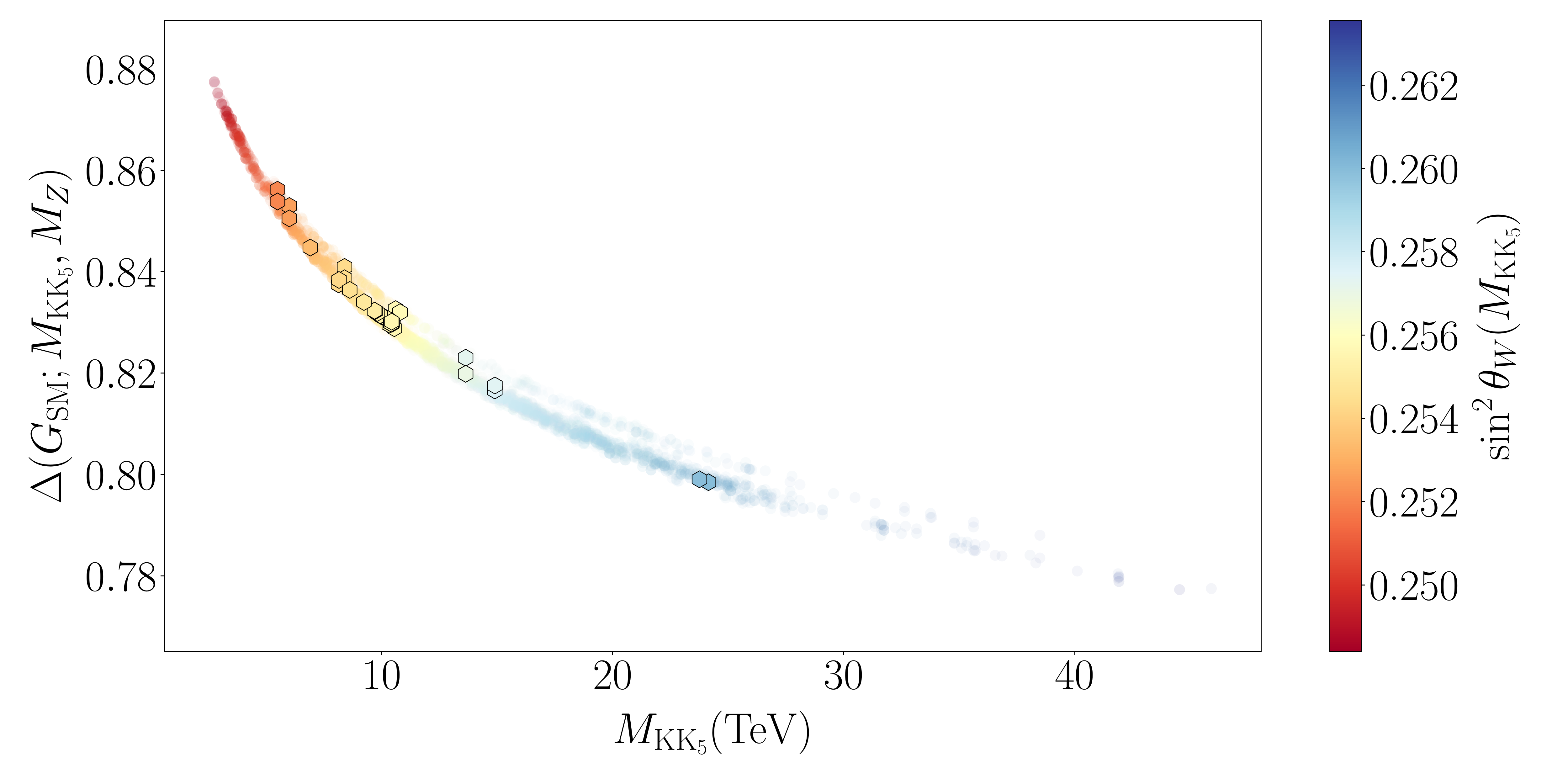}}
  \subfigure[\label{fig:PSabsDeV1Genb}]{\includegraphics[width=1.0\columnwidth]{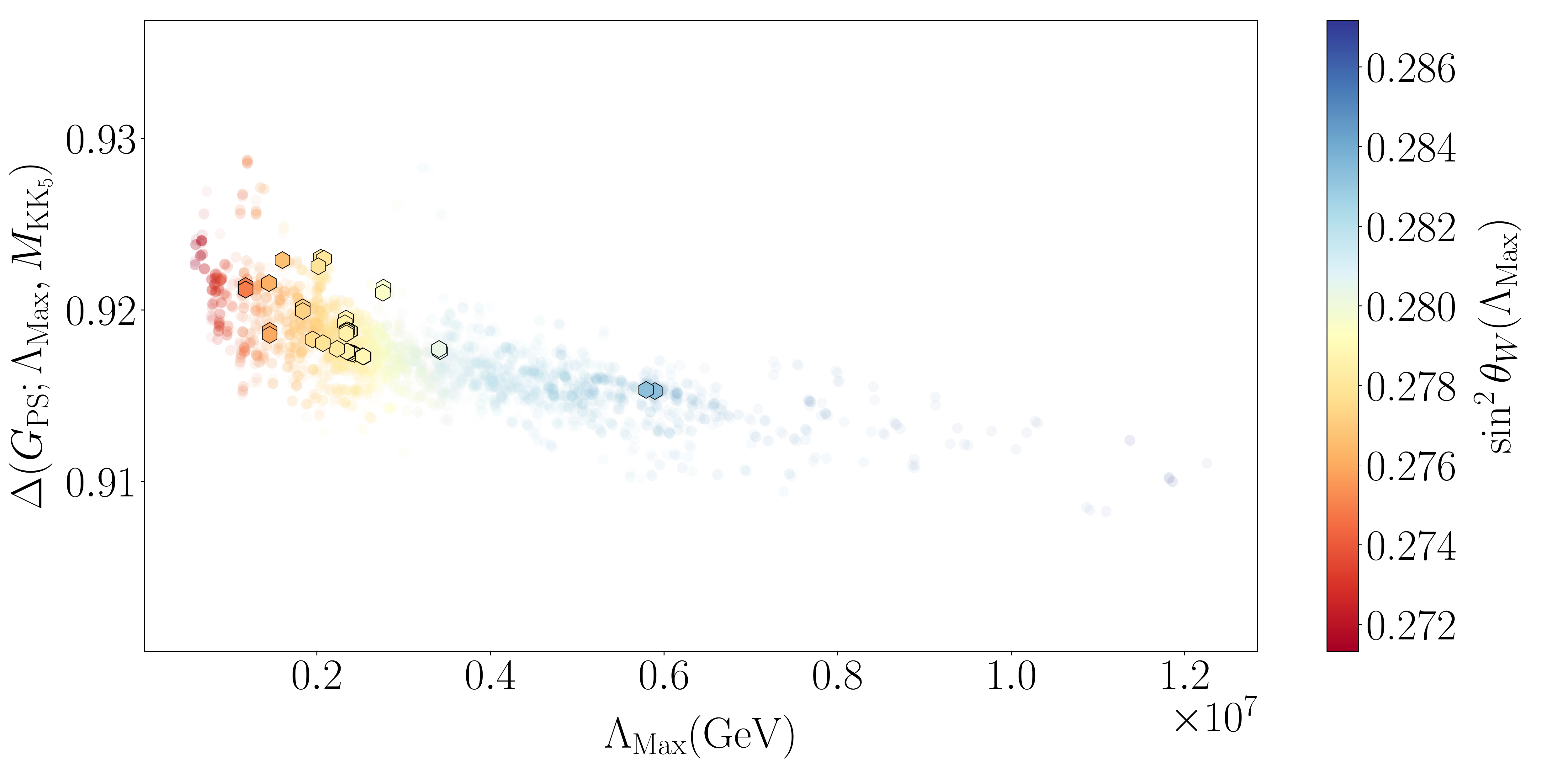}}
  \caption{(a) Scatter plot of the parameter space points for the $N_G=1$ case, where we use the same convention as in Fig.~\ref{fig:Res}. We now represent each point's value for the unification measure $\Delta(G_\text{SM}; M_{\text{KK}_5}, M_Z)$ in the 4D SM phase between the Kaluza-Klein scale $M_{\text{KK}_5}$, and $M_Z$, the respective KK scale, and the colour shading denotes the value of the Weinberg angle $\sin^2 \theta_W (M_{\text{KK}_5})$. (b) Correlation of the $N_G=1$ case in the 5D phase, $\Delta(G_\text{PS}; \Lambda_\text{Max}, M_{\text{KK}_5})$, shown as a function of the cut-off scale $\Lambda_\text{Max}$ where perturbativity is lost (see text for details). The colour shading again represents the Weinberg angle at the cut-off. Highlighted hexagon points refer to realistic low energy spectra compatible with exotics searches. \label{fig:SMabsDeV1Gen}
  }
 \end{center}
\end{figure*}

\begin{figure*}[!t]
 \begin{center}
 \subfigure[\label{fig:SMabsDeV3Gena}]{\includegraphics[width=1.0\columnwidth]{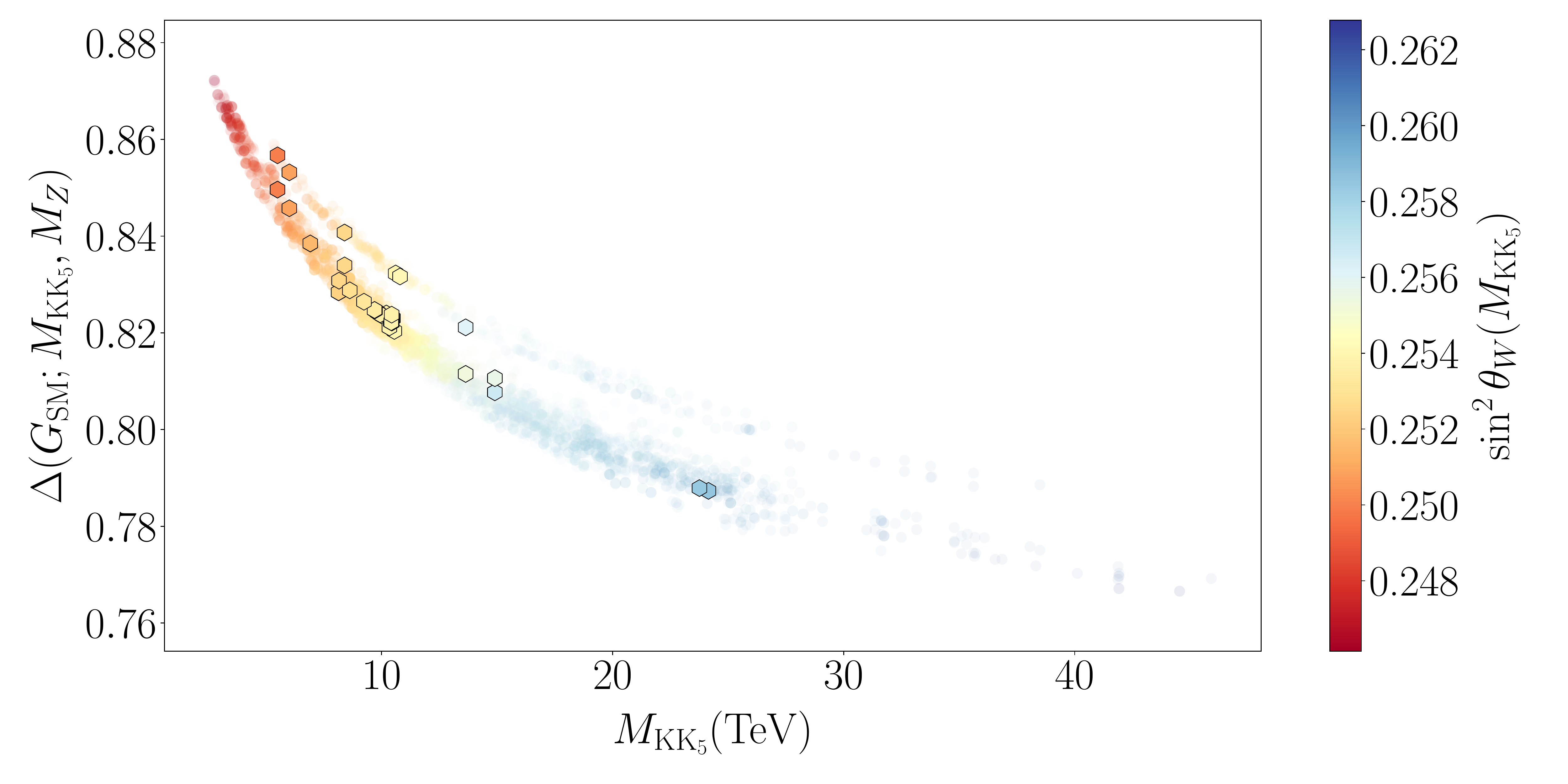}}
  \subfigure[\label{fig:PSabsDeV3Genb}]{\includegraphics[width=1.0\columnwidth]{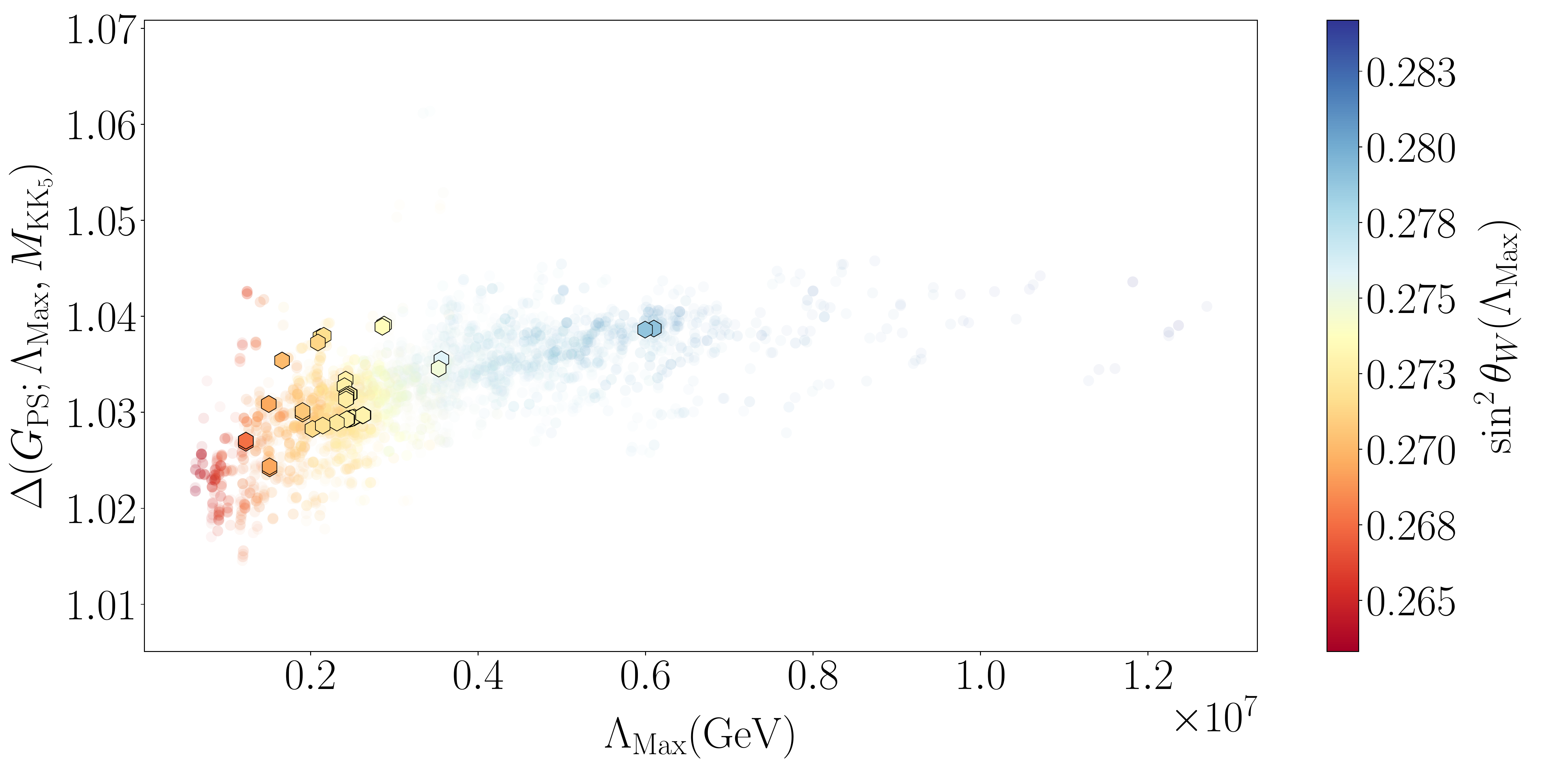}}
  \caption{Scatter plots analogous to Figs.~\ref{fig:SMabsDeV1Gena} and~\ref{fig:PSabsDeV1Genb} for the degenerate $N_G=3$ case. \label{fig:absDeV3Gen}
  }
 \end{center}
\end{figure*}

\section{Results and Conclusions}\label{sec:ResDiscConc}
The running is crucially influenced by the number of active fermion generations $N_G$. We will therefore
comment on our results for $N_G=1,3$ separately.

In the first case, we include only the third fermion generation as
mentioned before. This implicitly assumes that there is a large mass gap between the third family and the remaining two, decoupling the associated zero-mode KK states from the RGE flow (see Ref.~\cite{Hosotani:2017edv}). In the second case, we assume that all three SM generations are present and that different generational mass states are nearly degenerate. The comparison of these avenues contrasted with implications for unification can therefore act as a guideline for future model-building in the fermion sector.

To examine the extent to which the gauge couplings converge in the 4D, 5D regimes tensioned against the unification value of the Weinberg angle, we introduce a ``unification measure''
\begin{multline}
  \Delta(G; M_2, M_1) =
  \frac{\displaystyle \sum_{ i,j \in G | i \neq j} |\alpha_i (M_2) - \alpha_j(M_2)| }{ \displaystyle \sum_{ i,j \in G | i \neq j} |\alpha_i (M_1) - \alpha_j(M_1)| }\,,
\end{multline}
i.e. we consider the ratio of the sum of the mutual coupling deviations between two scales $M_2 > M_1$. $\alpha_i$ are the gauge group couplings of the subgroups that form the gauge group $G$.
This ratio measures how quickly the gauge couplings approach each other as a function of the energy scale. Since we are interested in gauge coupling unification at $M_2 > M_1$, values of $\Delta(G; M_2, M_1)$ refer to
\begin{equation}
  \Delta(G; M_2, M_1)
  \begin{cases}
    >1 \enskip \Leftrightarrow \enskip \text{departure from unification} \\
    <1 \enskip \Leftrightarrow \enskip \text{approaching unification} \\
    \sim 0 \enskip \Leftrightarrow \enskip \text{unification}
  \end{cases} \hspace{-0.2cm}.
\end{equation}
We plot this unification measure in the 4D SM phase between $M_Z, M_{\text{KK}_5}$, along with the Weinberg angle value at $M_{\text{KK}_5}$ in Figs.~\ref{fig:SMabsDeV1Gena}, \ref{fig:SMabsDeV3Gena} for the $N_G = 1$ and $N_G = 3$ cases.
Figs.~\ref{fig:PSabsDeV1Genb} and \ref{fig:PSabsDeV3Genb}. show the same measure for $N_G = 1$ and $N_G = 3$ in the 5D PS phase between $M_{\text{KK}_5}, \Lambda_\text{Max}$.

We start our discussion with the $N_G = 1$ case. Examining Fig.~\ref{fig:SMabsDeV1Gena} we can see that within the 4D SM phase, all the points that are consistent with the SM have a unification measure smaller than unity, where the evolved Weinberg angle is around $\sin^2\theta_W \simeq 0.25$.

\begin{figure*}[t!]%
    \centering
    \subfigure[~RGE evolution for the piecewise hypercharge coupling  $g_{1Y}$ for the sample point in Eqn.~\eqref{eqn:samplePoint} for the $N_G = 1$ case.]{{
    \includegraphics[width=0.96\columnwidth]{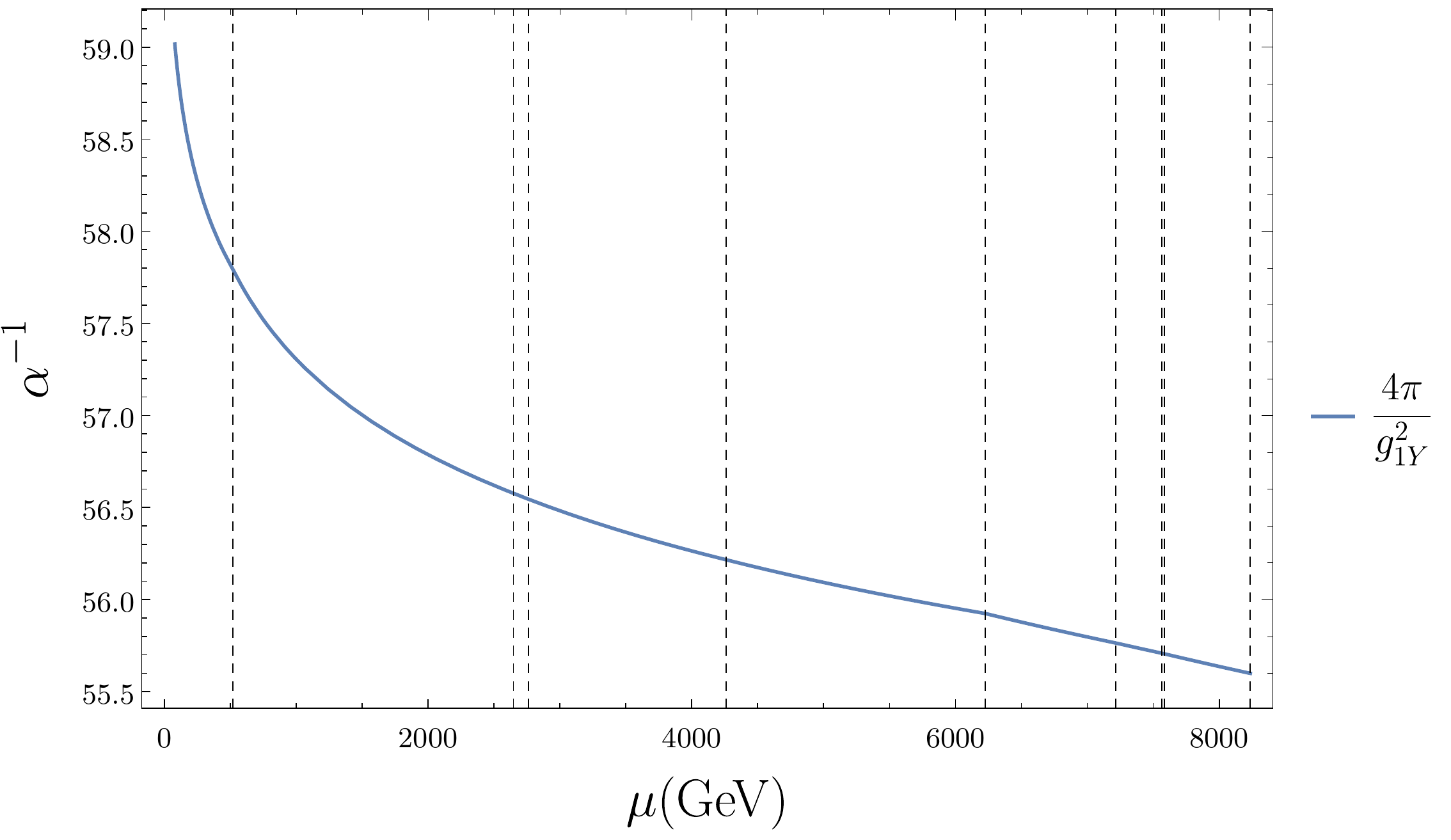}\label{fig:g1Y_1Gen}
    }}%
    %
    \subfigure[~RGE evolution for the piecewise hypercharge coupling  $g_{1Y}$ for the sample point in Eqn.~\eqref{eqn:samplePoint} for the $N_G = 3$ case.]{{
    \includegraphics[width=0.96\columnwidth]{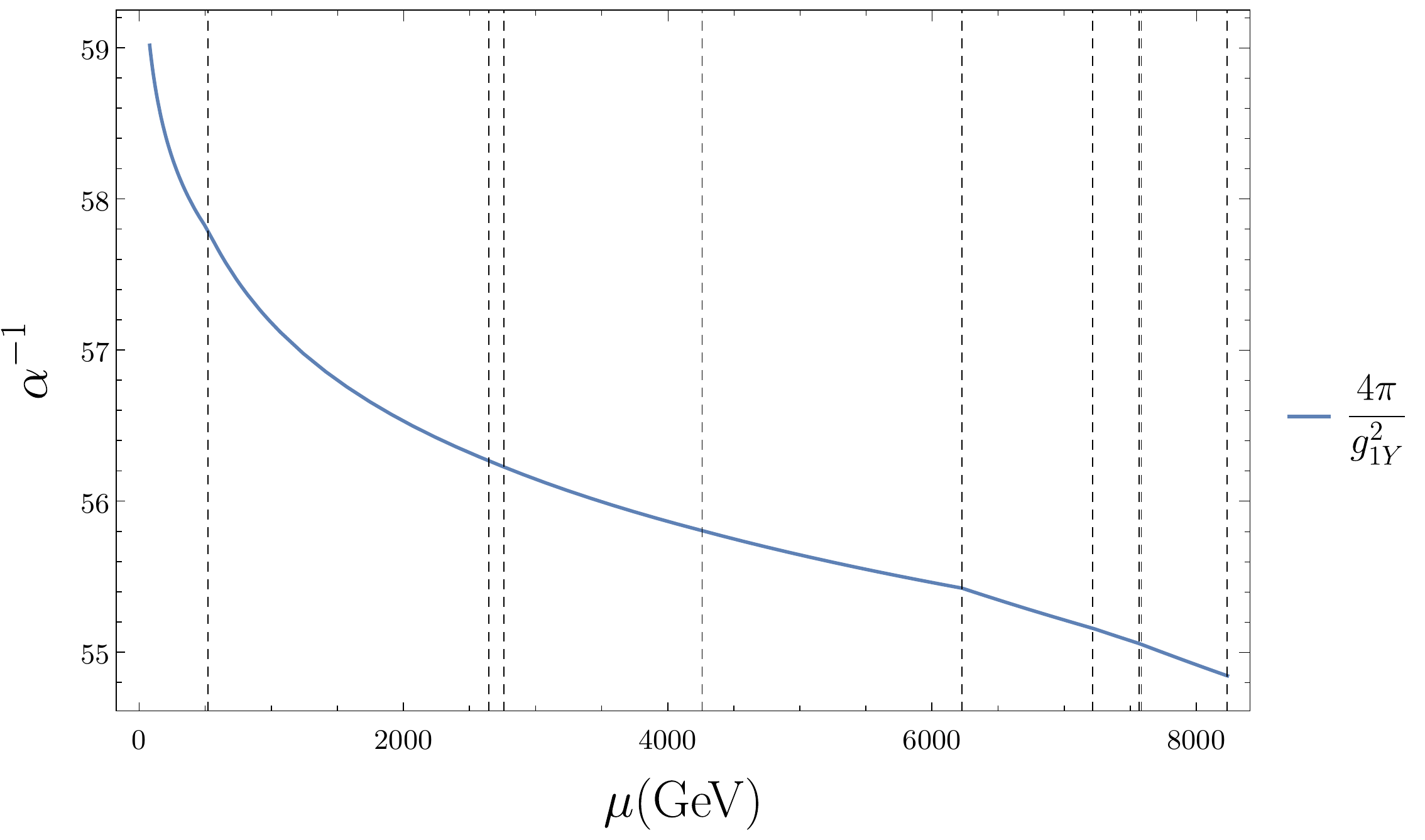}\label{fig:g1Y_3Gens}
     }}%
    \caption{Comparison between the piecewise RGE evolutions of the hypercharge couplings for the sample point in Eqn.~\eqref{eqn:samplePoint} between the $N_G=1$ and $N_G=3$ cases.}%
    \label{fig:g1YComparison}%
\end{figure*}
\begin{figure*}[t!]%
    \centering
    \subfigure[~RGE evolution for the piecewise weak coupling $g_{2L}$ for the sample point in Eqn.~\eqref{eqn:samplePoint} for the $N_G = 1$ case.]{{
    \includegraphics[width=0.96\columnwidth]{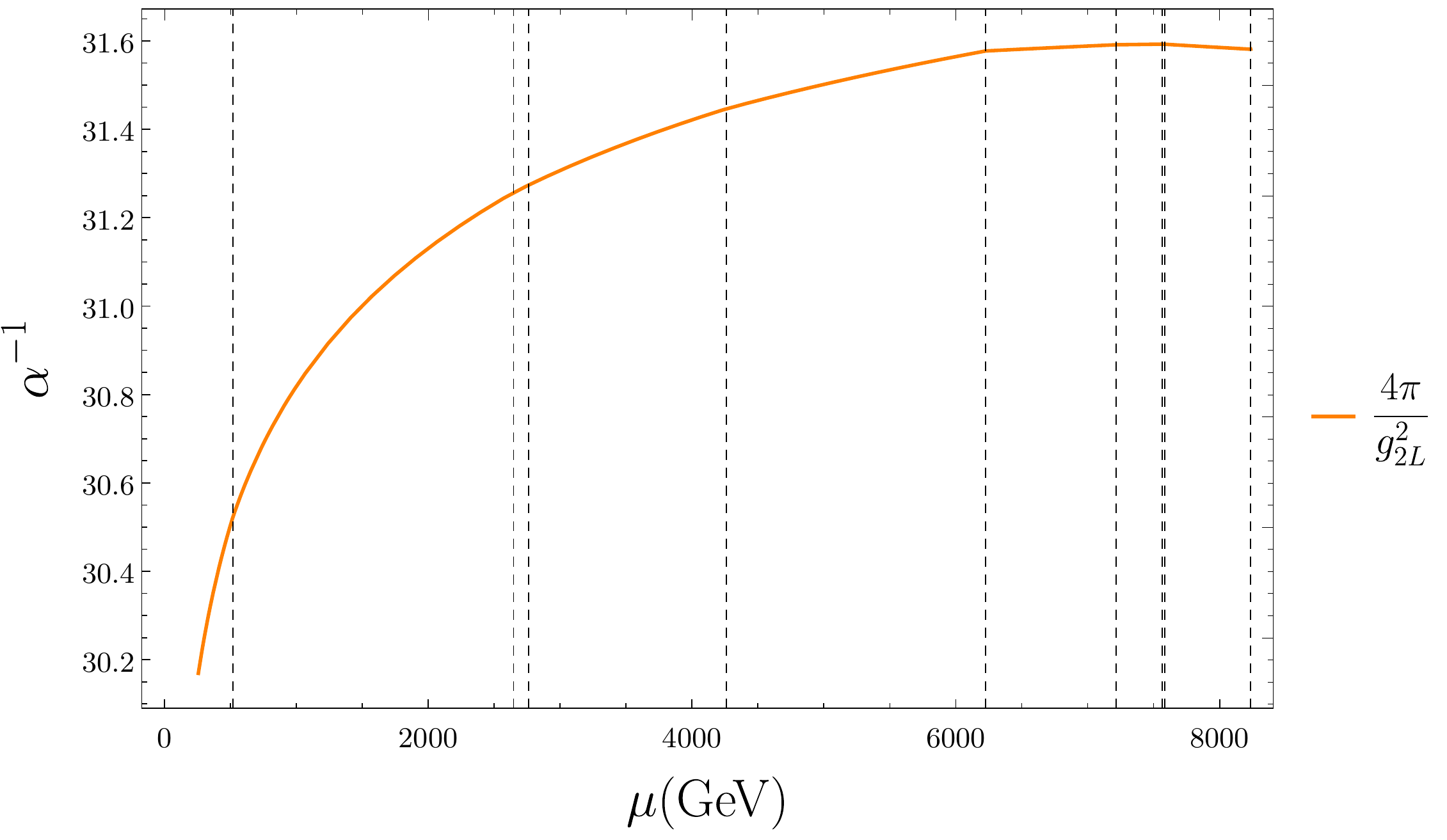}\label{fig:g2L_3Gens}
    }}%
    %
    \subfigure[~RGE evolution for the piecewise weak coupling $g_{2L}$ for the sample point in Eqn.~\eqref{eqn:samplePoint} for the $N_G = 3$ case.]{{
    \includegraphics[width=0.96\columnwidth]{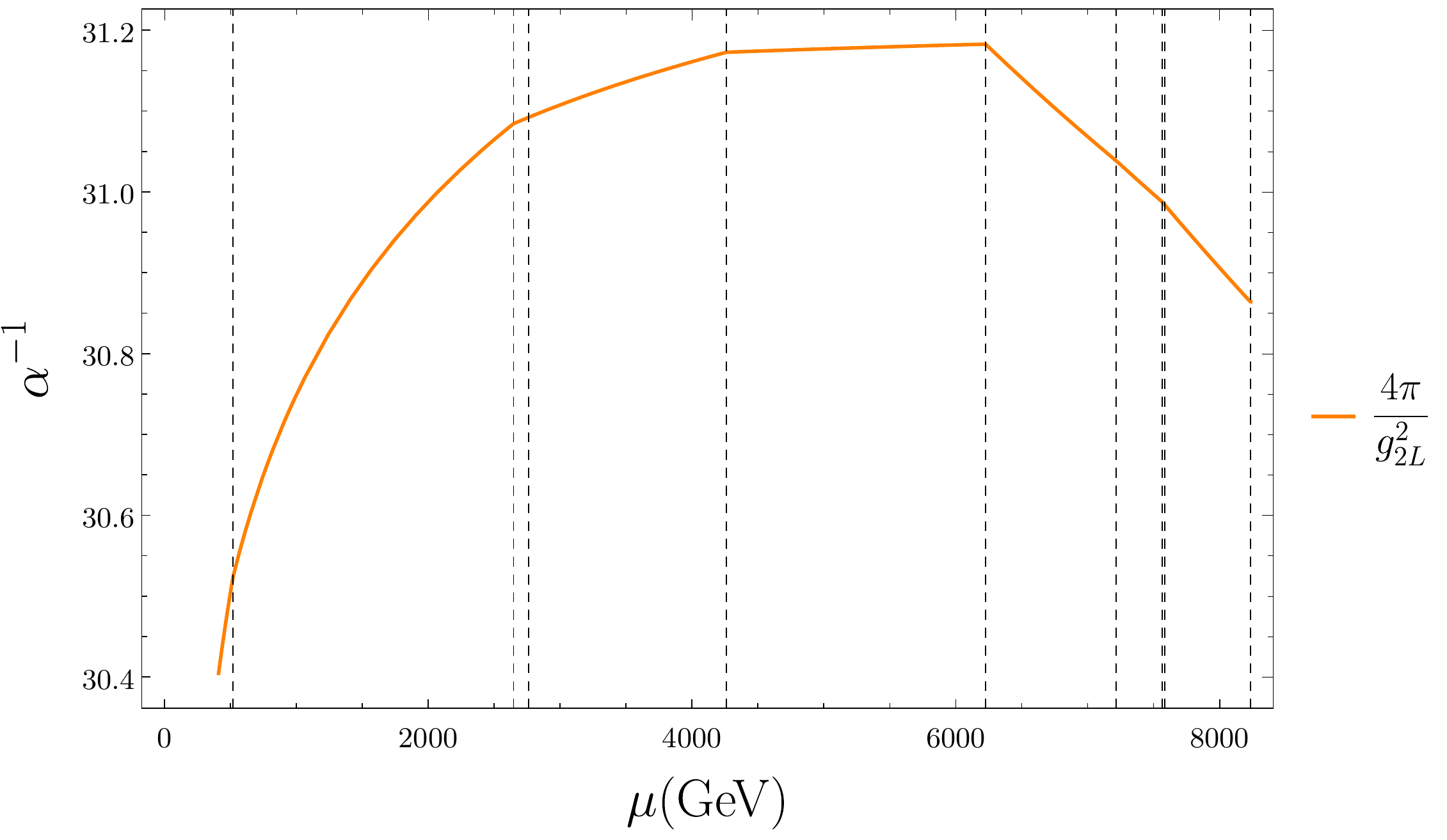}\label{fig:g2L_3Gens}
     }}%
    \caption{Comparison between the piecewise RGE evolutions of the weak couplings for the sample point in Eqn.~\eqref{eqn:samplePoint} between the $N_G=1$ and $N_G=3$ cases.}%
    \label{fig:g2LComparison}%
\end{figure*}

The Weinberg angle evolves towards its predicted unified value with a converging behaviour of the gauge couplings. The numerical results are similar between the $N_G = 1, 3$ cases, where in the $N_G = 3$ case, the unification measure is smaller due to the additional positive fermionic contributions which increase the slope of the running of the hypercharge coupling. We note that this effect is in competition with the weak corrections which tend to be strong enough to result in a change in the direction of the gauge coupling running away from asymptotic freedom. This in turn leads to a smaller Weinberg angle in the UV.
We can see this behaviour for the $N_G=1,3$ cases in Figs.~\ref{fig:g1YComparison} and \ref{fig:g2LComparison}.

In the 5D phase shown in Fig.~\ref{fig:PSabsDeV1Genb}, we see that the converging behaviour is maintained, where the unification measure increases compared to its 4D phase. The measure remains below uninty while the Weinberg angle also increases via the RGE flow.
This reflects the need for a complete set of RGEs to be performed within higher dimensional theories (see e.g. Refs.~\cite{Randall:2001gb, Randall:2001gc}).
Under the assumption that in the 6D phase of the theory the coupling behaviour remains similar,
we can infer that gauge coupling unification is consistent with the predicted value for the Weinberg angle. Put differently, the cut-off scale depicted in Fig.~\ref{fig:PSabsDeV1Genb} provides us with a lower bound for the unification scale  $M_\text{GUT} > \Lambda_\text{Max}$ which is dictated by gauge coupling unification and consistency with the Weinberg angle prediction.

Let us turn to the $N_G=3$ case, where we observe an amplified behaviour of the aforementioned effect of the KK states (Fig.~\ref{fig:SMabsDeV3Gena}) due to their increased number.
In total, this leads to gauge couplings getting increasingly pushed away from unification in the 5D phase, while the Weinberg angle flow is still consistent with its unification value.
Under the assumption, that this behaviour continues in the 6D theory, we could face a potential inconsistency arising from reaching the predicted $SU(5)$ value for the Weinberg angle, but not achieving gauge coupling unification.
While this could be compensated by large radiative corrections that shift the Weinberg angle away from the GUT hypothesis, this sets fairly tight constraints on the dynamics of the fermion sector.

We finally comment on the impact of uncertainties, in particular uncertainties of the input parameters $\alpha_{3C}$ and $\sin^2 \theta_W$ at the weak scale. Errors as small as $\sigma({\alpha_{3C}}) = \pm 0.00074$ are possible from a theoretical perspective (e.g. Ref.~\cite{Chakraborty:2014aca}), and we consider a conservative $5\%$ uncertainty in the value of the Weinberg angle where $\sigma({\sin^2 \theta_W}) = \pm 0.01156$ for demonstration purposes. Taking into account both of these uncertainties, we perform our analysis for the sample point highlighted in Eq.~\eqref{eqn:samplePoint}.

In both the $N_G = 1$ and $N_G = 3$ cases the percentage difference arising in the unification measure at $M_{\mathrm{KK}_5}$ amounts to $\sim 2\%$.
 This effect is less pronounced at $\Lambda_{\mathrm{Max}}$, decreasing to $\sim 0.2 \%$ for $N_G = 1$ and $\sim 0.1 \%$ for $N_G = 3$.
In the $N_G = 1$ case the Weinberg angle at $M_{\mathrm{KK}_5}$ is affected by $\sim 4.7 \%$, and decreases at $\Lambda_{\mathrm{Max}}$ to $\sim 3.9 \%$.
In the $N_G = 3$ case the impact on the Weinberg angle is similar; at $M_{\mathrm{KK}_5}$ we obtain $\sim 4.8\%$, and at $\Lambda_{\mathrm{Max}}$ we have an increase to $\sim 4.92\%$.

\section{Conclusions}
\label{sec:Conc}
Grand Unified Theories are attractive solutions to shortcomings of the Standard Model of Particle Physics. In non-supersymmetric realisations, scale separations can be achieved by employing higher dimensional background geometry~\cite{Randall:1999ee}, where electroweak symmetry breaking can also be implemented elegantly as a radiative phenomenon~\cite{Hosotani:1983xw}. Transitioning through the different phases of such scenarios is less straightforward compared to applications in ``standard'' 4D GUTs (see e.g.~\cite{Babu:2015bna,Lazarides:1980nt,Hall:1993gn,Barr:1981qv,Dimopoulos:1981zb,Derendinger:1983aj,Antoniadis:1987dx,Maekawa:2002bk}).

This is the purpose of our study: a detailed analysis of the 4D and 5D phases of the model of~\cite{Hosotani:2017edv,Hosotani:2017hmu}, contrasted with electroweak scale measurements as well as LHC constraints. We pay particular attention to the Weinberg angle, whose size is determined by $SU(5)$ relations, and can therefore be used to test gauge unification (or lack thereof). While a fully conclusive test will need a full investigation of the 6D phase of the theory, which we leave for future work, we gather evidence that the 4D and 5D effective theories can remain under perturbative control up to scales of $\sim 10^7$~GeV. If unification is to be approached in a controlled way, new dynamics should appear at scales about two orders of magnitude above the 5D compactification scale. This scale can be interpreted as a lower limit on the GUT scale $\sim 5000$ TeV in the light of observed physics at and around the electroweak scale.

Fermionic thresholds crucially impact the running of couplings and as a consequence, the model-building aspects related to the three fermion generations plays an important role in the high energy behaviour of the theory. Unless there is a hierarchical approach to lifting the zero modes of the fermion fields to their observed SM values, the 6D theory will play a more important role in achieving unification in the sense of Eq.~\eqref{eq:weinuv}.

\acknowledgements
We thank the referee of Physics Letters B for their comments.

CE acknowledges support by the Durham Institute for Particle Physics Phenomenology (IPPP) Associateship Scheme and thanks the IPPP for the hospitality extended to him while this work was finalised. CE and DJM are supported by the UK Science and Technology Facilities Council (STFC) under grant ST/P000746/1. DDS is supported by a University of Glasgow College of Science \& Engineering PhD scholarship.

\appendix
\section*{Appendix}
\subsection{5D RGE contributions}\label{appendix:5DRGEs}
The form of the $SU(N)$  corrections $\overline{\Delta}_N (\xi)$ for a generic field $\xi$ are specified in \cite{Choi:2002ps}.
Starting with $SU(4)_C$, the corrections due to scalars, gauge fields and fermions are,
\begin{align*}
 \overline{\Delta}_{4C} & (\phi) =  2  \overline{\Delta}_{4C}^{--} (\mathbf{15}) \,,\\
 \overline{\Delta}_{4C} & (\psi) =   \\
    & \phantom{+ } 3 \left( \overline{\Delta}_{4C}^{++} (\mathbf{4}_L) +  \overline{\Delta}_{4C}^{++} (\mathbf{4}_R) +  \overline{\Delta}_{4C}^{--} (\mathbf{4}_L) +  \overline{\Delta}_{4C}^{--} (\mathbf{4}_R) \right) \\
                             & \hspace{-0.5em} + 3  \left( \overline{\Delta}_{4C}^{++} (\mathbf{6}_L)+ \overline{\Delta}_{4C}^{++} (\mathbf{6}_R) + \overline{\Delta}_{4C}^{--} (\mathbf{6}_L) + \overline{\Delta}_{4C}^{--} (\mathbf{6}_R) \right)  \\
 & \hspace{-0.5em} + \overline{\Delta}_{4C}^{+-} (\mathbf{4}_L) + \overline{\Delta}_{4C}^{-+} (\mathbf{4}_R) + \overline{\Delta}_{4C}^{-+} (\mathbf{4}_L) + \overline{\Delta}_{4C}^{+-} (\mathbf{4}_R) \,,\\
 \overline{\Delta}_{4C} & (A)   =  \overline{\Delta}_{3C}^{++} (\mathbf{8}) + \overline{\Delta}_{3C}^{-+} (\mathbf{3}) + \overline{\Delta}_{3C}^{-+} (\overline{\mathbf{3}} )\,,                                                                &   &
\end{align*}
where the $\pm$ signs refer to the parity assignments, the factor of 2 arises from the $A_{y,w}$ components, and the factors of 3 arise from generational indices $\alpha=1, 2, 3$ and $\beta=1, 2, 3$.

The contribution of the gauge fields is obtained by decomposing the adjoint $\mathbf{15}$ of $SU(4)$ under the breaking chain $SU(4) \rightarrow SU(3) \times U(1)$, and adding each subcomponent's contribution separately based on their parity assignment. Concretely, $\mathbf{15} \rightarrow (\mathbf{8}, 0) \oplus (\mathbf{3}, +4/3) \oplus (\overline{\mathbf{3}}, -4/3) \oplus (1, 0)$. For the singlet representation we have $\Delta_3(1) = 0$. Given that we effectively deal with a symmetry projection $SU(5) \cap G_\text{PS}  = G_\text{SM}$,
we treat these multiplets separately due to their different effective parity assignments, see Tabs.~\ref{table:GaugeParity} and~\ref{table:GaugeScalarParity}.

Moving on to $SU(2)_L$, the corrections are
\begin{align*}
 \overline{\Delta}_{2L}& (\phi)  = \overline{\Delta}_{2L}^{++} (\mathbf{2}) +  2  \overline{\Delta}_{2L}^{--} (\mathbf{3})   \,,\\
 \overline{\Delta}_{2L}& (\psi)  =  3  \left( \overline{\Delta}_{2L}^{++} (\mathbf{2}_L) +  \overline{\Delta}_{2L}^{--} (\mathbf{2}_R)  \right) \\
                                 & \hspace{-0.25em}+ 3 \left( \overline{\Delta}_{2L}^{++} (\mathbf{2}_L) + \overline{\Delta}_{2L}^{++} (\mathbf{2}_R) + \overline{\Delta}_{2L}^{--} (\mathbf{2}_L) + \overline{\Delta}_{2L}^{--} (\mathbf{2}_R)  \right) \\
 & \hspace{-0.5em} + \left( \overline{\Delta}_{2L}^{+-} (\mathbf{2}_L) + \overline{\Delta}_{2L}^{-+} (\mathbf{2}_R) \right) \,,\\
 \overline{\Delta}_{2L} & (A)   = \overline{\Delta}_{2L}^{++} (\mathbf{3}) + \overline{\Delta}_{2L}^{--} (\mathbf{2})\,,                                                                                                           &   &
\end{align*}
where the factors of 3 arise from $\alpha = 1, 2, 3$ and $\beta = 1, 2, 3$. $SU(2)_R$ has almost identical corrections apart from those originating from $\Psi^4_\mathbf{32}$, where $(+,-)$ and $(-,+)$ are swapped for R, L indices, and the gauge contribution,
\begin{align*}
 \overline{\Delta}_{2R}& (\phi) =  \hspace{0.2cm}\overline{\Delta}_{2R}^{++} (\mathbf{2}) +  2  \overline{\Delta}_{2R}^{--} (\mathbf{3})\,,  \\
 \overline{\Delta}_{2R}& (\psi) =  \hspace{0.2cm}  3  \left( \overline{\Delta}_{2R}^{++} (\mathbf{2}_L) +  \overline{\Delta}_{2R}^{--} (\mathbf{2}_R)  \right) \\
                                 & \hspace{-0.5em} + 3 \left( \overline{\Delta}_{2R}^{++} (\mathbf{2}_L) + \overline{\Delta}_{2R}^{++} (\mathbf{2}_R) + \overline{\Delta}_{2R}^{--} (\mathbf{2}_L) + \overline{\Delta}_{2R}^{--} (\mathbf{2}_R)  \right)  \\
 & \hspace{-0.5em} + \left( \overline{\Delta}_{2R}^{-+} (\mathbf{2}_L) + \overline{\Delta}_{2R}^{+-} (\mathbf{2}_R) \right)\,, \\
 \overline{\Delta}_{2R} & (A) = \hspace{0.2cm} \overline{\Delta}_{2R}^{-+} (\mathbf{3}) + \overline{\Delta}_{2R}^{--} (\mathbf{2}).                                                                                                           &   &
\end{align*}

The explicit form of the corrections are listed below.
\begin{widetext}
 \begin{itemize}
  \item \textbf{5D scalars: $\phi(x^\mu, y)$} have a contribution to the gauge coupling $g_a$, corresponding to $SU(N^a)$, of  the form:
        \begin{align*}
         \overline{\Delta}_a (\phi ; \mu) = \frac{1}{12}
           & \Bigg[ T_a(\phi_{++}) \left[ \ln \left( \frac{\Lambda}{k} \right) - 3 \int_0^1 du\, F(u) \ln N_{\phi_{++}} \left( \frac{i u }{2} \sqrt{\mu^2} \right)  \right]   \\
           & - 3 T_a (\phi_{+-}) \int_0^1 du\, F(u) \ln N_{\phi_{+-}} \left( \frac{i u }{2} \sqrt{\mu^2} \right) -                                                             \\
           & - 3 T_a (\phi_{-+}) \int_0^1 du \,F(u) \ln N_{\phi_{-+}} \left( \frac{i u }{2} \sqrt{\mu^2} \right) -                                                             \\
           & - T_a(\phi_{--}) \left[ \ln \left( \frac{\Lambda}{k} \right) + 3 \int_0^1 du\, F(u) \ln N_{\phi_{--}} \left( \frac{i u }{2} {\sqrt{\mu^2}} \right)  \right] \Bigg]\,,
        \end{align*}
        where $T_a(\phi)$ is the Dynkin index of the $SU(N^a)$ representation for $\phi$, $F(u) = u(1 - u^2)^{\frac{1}{2}}$, and  $N_{\phi_{\pm, \pm}}$ are the $N-$functions from Appendix~\ref{apdx:ScanAppendix} with,
        \begin{equation}
         (Z_\phi, Z'_\phi, \{ \mathcal{P}_\phi\}) = (\pm, \pm, 4, 0, 0, 2),
        \end{equation}
        where parameter set $\{\mathcal{P}_\phi\}$ is defined in Appendix~\ref{apdx:ScanAppendix}.
  \item \textbf{5D fermion fields $\psi(x^\mu, y)$} have a contribution to the gauge coupling $g_a$, corresponding to $SU(N^a)$, of  the form:
        \begin{align*}
         \overline{\Delta}_a (\psi ; \mu) = \frac{1}{3}
           & \Bigg[ T_a(\psi_{++}) \left[ 2\ln \left( \frac{k}{\mu} \right) - k L_5 + 3 \int_0^1 du\, G(u) \ln N_{\psi_{++}} \left( \frac{i u }{2} \sqrt{\mu^2} \right)  \right]   \\
           & + T_a (\psi_{+-})\left[ -k L_5 + 3 \int_0^1 du\, G(u) \ln N_{\psi_{+-}} \left( \frac{i u }{2} \sqrt{\mu^2} \right) \right]                                             \\
           & + T_a (\psi_{-+})\left[ k L_5 + 3 \int_0^1 du\, G(u) \ln N_{\psi_{-+}} \left( \frac{i u }{2} \sqrt{\mu^2} \right) \right]                                               \\
           & + T_a(\psi_{--}) \left[ 2\ln \left( \frac{k}{\mu} \right) - k L_5 + 3 \int_0^1 du\, G(u) \ln N_{\psi_{--}} \left( \frac{i u }{2} \sqrt{\mu^2} \right)  \right] \Bigg]\,,
        \end{align*}
        where $T_a(\psi)$ is the Dynkin index of the $SU(N^a)$ representation for $\psi$, \mbox{$G(u) = u(1 - u^2)^{\frac{1}{2}} - u(1-u^2)^{-\frac{1}{2}}$}, and $N_{\psi_{\pm, \pm}}$ are the $N-$functions from Appendix~\ref{apdx:ScanAppendix}, where
        \begin{equation}
         (Z_\phi, Z'_\phi, \{ \mathcal{P}_\phi\}) =
         \begin{cases}
          (-, -,\{ 1, +c, +c, |c - \frac{1}{2}|\}) \quad \text{for} \quad N_{\psi_{++}} \\
          (-, -,\{ 1, -c, -c, |c + \frac{1}{2}|\}) \quad \text{for} \quad N_{\psi_{--}} \\
          (+, -,\{ 1, -c, -c, |c + \frac{1}{2}|\}) \quad \text{for} \quad N_{\psi_{+-}} \\
          (-, +,\{ 1, -c, -c, |c + \frac{1}{2}|\}) \quad \text{for} \quad N_{\psi_{-+}} \\
         \end{cases}.
        \end{equation}

  \item \textbf{5D Gauge fields $A_M (x^\mu, y)$} have a contribution to the gauge coupling $g_a$, corresponding to $SU(N^a)$, of  the form:
        \begin{align*}
         \overline{\Delta}_a (A ; \mu) = \frac{1}{12}
           & \Bigg[ T_a(A_{++}) \left[ 23 \ln \left( \frac{\mu}{\Lambda} \right) + 21 \ln \left( \frac{\mu}{k} \right) + 22 k L_5 +  \int_0^1 du\, K(u) \ln N_{A_{++}} \left( \frac{i u}{2} \sqrt{\mu^2} \right)  \right]   \\
           & + T_a(A_{+-}) \left[ -k L_5  + \int_0^1 du\, K(u) \ln N_{A_{+-}} \left( \frac{i u}{2} \sqrt{\mu^2} \right)   \right]                                                                                            \\
           & + T_a(A_{-+}) \left[ k L_5  + \int_0^1 du \,K(u) \ln N_{A_{-+}} \left( \frac{i u}{2} \sqrt{\mu^2} \right)   \right]                                                                                             \\
           & + T_a(A_{--}) \left[ 23 \ln \left( \frac{\Lambda}{k} \right) + 2 \ln \left( \frac{k}{\mu} \right) - k L_5 +  \int_0^1 du\, K(u) \ln N_{A_{--}} \left( \frac{i u}{2} \sqrt{\mu^2} \right)  \right]  \Bigg]\,,
        \end{align*}
        where $T_a(A)$ is the Dynkin index of the $SU(N^a)$ representation for $A$, $K(u) = -9u(1 - u^2)^{\frac{1}{2}} + 24 u(1-u^2)^{-\frac{1}{2}}$, and $N_{\psi_{\pm, \pm}}$ are the $N-$functions from Appendix.~\ref{apdx:ScanAppendix}, where
\begin{equation}
         (Z_\phi, Z'_\phi, \{ \mathcal{P}_\phi\}) =
         \begin{cases}
          (-, -,\{ 4, 2, 2, 0\}) \quad \text{for} \quad N_{A_{++}} \\
          (-, -,\{ 2, 0, 0, 1\}) \quad \text{for} \quad N_{A_{--}} \\
          (+, -,\{ 2, 0, 0, 1\}) \quad \text{for} \quad N_{A_{+-}} \\
          (-, +,\{ 2, 0, 0, 1\}) \quad \text{for} \quad N_{A_{-+}}
         \end{cases}\hspace{-0.2cm}.
        \end{equation}
\end{itemize}

\subsection{$N_{(\pm, \pm)}(\mu)$ functions}\label{apdx:ScanAppendix}
The $N-$functions for a generic field $\xi$ where $\xi \in \left\{ A_\mu, \phi, e^{-2 k L_5 |y|} \psi_L, e^{-2 k L_5 |y|} \psi_R \right\}$ with $(\mathds{Z}_2 , \mathds{Z}'_2)$  parity assignments depend on the renormalisation scale $\mu$, the AdS curvature $k$, the warp factor $z_L = \exp( k L_5)$, the $\xi$ field set of defining parameters
$\{\mathcal{P}_\xi\} = \{  s_\xi , (r_0)_\xi, (r_\pi)_\xi, \alpha \}$, where $s$ is associated with the spin of the field,
  \begin{equation}
   s_\xi= \{2, 4, 1, 1,  \} \quad \text{for} \quad  \xi \in \left\{ A_\mu, \phi, e^{-2 k L_5 |y|} \psi_L, e^{-2 k L_5 |y|} \psi_R \right\},
  \end{equation}
  and is related to $\alpha$ as
  \begin{equation}
   \alpha = \sqrt{ \left(\frac{s}{2}\right)^2 + M^2_\xi } \quad \text{where} \quad  M^2_\xi \in \left\{ 0, 0, c(c+1), c(c-1) \right\} .
  \end{equation}
  Note that the model explored in this paper does not have any bulk masses present for the gauge fields.
  The closed form for the $N-$ functions is given by
  \begin{align*}
   N_{\xi_{(+,+)}}(\mu ; \{\mathcal{P}_\xi\}) = & - \Bigg[ \left[ \frac{s_\xi}{2} - (r_0)_\xi \right] J_\alpha \left( \frac{\mu}{k} \right) + \frac{\mu}{k} J'_\alpha \left( \frac{\mu}{k} \right) \Bigg]
   \Bigg[ \left[ \frac{s_\xi}{2} - (r_\pi)_\xi \right] Y_\alpha \left( \frac{\mu}{k z_L} \right) + \frac{\mu}{k z_L} Y'_\alpha \left( \frac{\mu}{k z_L} \right) \Bigg]  \\
                                                & + \Bigg[ \left[ \frac{s_\xi}{2} - (r_\pi)_\xi \right] J_\alpha \left( \frac{\mu}{k z_L} \right) + \frac{\mu}{k z_L} J'_\alpha \left( \frac{\mu}{k z_L} \right) \Bigg]
   \Bigg[ \left[ \frac{s_\xi}{2} - (r_0)_\xi \right] Y_\alpha \left( \frac{\mu}{k} \right) + \frac{\mu}{k} Y'_\alpha \left( \frac{\mu}{k} \right) \Bigg]\, ,
  \end{align*}
  \begin{align*}
   N_{\xi_{(+,-)}}(\mu; \{\mathcal{P}_\xi\}) = & - Y_\alpha \left( \frac{\mu}{k z_L} \right) \Bigg[ \left[ \frac{s_\xi}{2} - (r_0)_\xi \right]  J_\alpha \left( \frac{\mu}{k} \right) + \frac{\mu}{k} J'_\alpha \left( \frac{\mu}{k} \right) \Bigg]  \\
                                               & + J_\alpha \left( \frac{\mu}{k z_L} \right) \Bigg[ \left[ \frac{s_\xi}{2} - (r_0)_\xi \right] Y_\alpha \left( \frac{\mu}{k} \right) + \frac{\mu}{k} Y'_\alpha \left( \frac{\mu}{k} \right) \Bigg]\,,
  \end{align*}
  \begin{align*}
   N_{\xi_{(-,+)}}(\mu; \{\mathcal{P}_\xi\}) = & + J_\alpha \left( \frac{\mu}{k} \right) \Bigg[ \left[ \frac{s_\xi}{2} - (r_\pi)_\xi \right]  Y_\alpha \left( \frac{\mu}{k z_L} \right) + \frac{\mu}{k zL} Y'_\alpha \left( \frac{\mu}{k zL} \right) \Bigg]  \\
                                               & - Y_\alpha \left( \frac{\mu}{k } \right) \Bigg[ \left[ \frac{s_\xi}{2} - (r_\pi)_\xi \right] J_\alpha \left( \frac{\mu}{k zL} \right) + \frac{\mu}{k zL} J'_\alpha \left( \frac{\mu}{k zL} \right) \Bigg]\,,
  \end{align*}
  \begin{equation}
   N_{\xi_{(-,-)}}(\mu; \{\mathcal{P}_\xi\}) =  J_\alpha \left( \frac{\mu}{k} \right) Y_\alpha \left( \frac{\mu}{k z_L} \right) - J_\alpha \left( \frac{\mu}{k z_L} \right) Y_\alpha \left( \frac{\mu}{k} \right)\, ,
  \end{equation}
  where $(r_0)_\xi, (r_\pi)_\xi$ denote the 5D mass parameters at the branes. In our case, they take the simplified form for $\xi \in \left\{ A_\mu, \phi, e^{-2 k L_5 |y|} \psi_L, e^{-2 k L_5 |y|} \psi_R \right\}$, of
\end{widetext}
  \begin{equation}
   (r_0)_\xi =  (r_\pi)_\xi = \{ 0, 0, -c_\xi, c_\xi\}\,,
  \end{equation}
  where $c_\xi$ corresponds to the parent field's original 5D mass parameter $c_\xi \in \{c_0, c_1, c_2, c'_0\}$. Note that we do not have any artificially introduced brane masses for the scalar fields in the 5D limit.

 \bibliography{paper.bbl}

\end{document}